
\input phyzzx.tex
%
%

   \message{V 1.23 mods and bug fixes by M.Weinstein}
   \def\unlock{\catcode`@=11}

   \def\lock{\catcode`@=12}

   \unlock
%
%
   \def\PRrefmark#1{\unskip~[#1]}
   \def\refitem#1{\ifPhysRev\r@fitem{[#1]}\else\r@fitem{#1.}\fi}
   \def\generatefootsymbol{%
      \ifcase\footsymbolcount\fd@f 13F \or \fd@f 279 \or \fd@f 27A
          \or \fd@f 278 \or \fd@f 27B
      \else%
         \ifnum\footsymbolcount <0%
            \xdef\footsymbol{\number-\footsymbolcount}%
         \else%
            \fd@f 203
               {\loop \ifnum\footsymbolcount >5
                  \fd@f{203 \footsymbol }
                  \advance\footsymbolcount by -1
                \repeat
               }
         \fi%
      \fi%
   }
   \def\OldPhysRevRefmark{\let\PRrefmark=\attach}
   \def\OldPRRefitem#1{\r@fitem{#1.}}
   \def\OldPhysRevRefitem{\let\refitem=\OldPRRefitem}
   \def\NPrefs{\let\refmark=\NPrefmark \let\refitem=\NPrefitem}
%
    \newif\iffileexists              \fileexistsfalse
    \newif\ifforwardrefson           \forwardrefsontrue
    \newif\ifamiga                   \amigafalse
    \newif\iflinkedinput             \linkedinputtrue
    \newif\iflinkopen                \linkopenfalse
    \newif\ifcsnameopen              \csnameopenfalse
    \newif\ifdummypictures           \dummypicturesfalse
    \newif\ifcontentson              \contentsonfalse
    \newif\ifcontentsopen            \contentsopenfalse
    \newif\ifmakename                \makenamefalse
    \newif\ifverbdone
    \newif\ifusechapterlabel         \usechapterlabelfalse
    \newif\ifstartofchapter          \startofchapterfalse
    \newif\iftableofplates           \tableofplatesfalse
    \newif\ifplatesopen              \platesopenfalse
    \newif\iftableoftables           \tableoftablesfalse
    \newif\iftableoftablesopen       \tableoftablesopenfalse
    \newif\ifwarncsname              \warncsnamefalse
%
    \newwrite\linkwrite
    \newwrite\csnamewrite
    \newwrite\contentswrite
    \newwrite\plateswrite
    \newwrite\tableoftableswrite
    \newread\testifexists
    \newread\verbinfile

    \newtoks\jobdir                  \jobdir={}
    \newtoks\tempnametoks            \tempnametoks={}
    \newtoks\oldheadline             \oldheadline={}
    \newtoks\oldfootline             \oldfootline={}
    \newtoks\subsectstyle            \subsectstyle={\Number}
    \newtoks\subsubsectstyle         \subsubsectstyle={\Number}
    \newtoks\runningheadlines        \runningheadlines={\relax}
    \newtoks\chapterformat           \chapterformat={\titlestyle}
    \newtoks\sectionformat           \sectionformat={\relax}
    \newtoks\subsectionformat        \subsectionformat={\relax}
    \newtoks\subsubsectionformat     \subsubsectionformat={\relax}
    \newtoks\chapterfontstyle        \chapterfontstyle={\bf}
    \newtoks\sectionfontstyle        \sectionfontstyle={\rm}
    \newtoks\subsectionfontstyle     \subsectionfontstyle={\rm}
    \newtoks\sectionfontstyleb       \sectionfontstyleb={\caps}
    \newtoks\subsubsectionfontstyle  \subsubsectionfontstyle={\rm}

    \newcount\subsectnumber           \subsectnumber=0
    \newcount\subsubsectnumber        \subsubsectnumber=0


   \newdimen\pictureindent           \pictureindent=15pt
   \newdimen\str
   \newdimen\squareht
   \newdimen\squarewd
   \newskip\doublecolskip
   \newskip\tableoftablesskip        \tableoftablesskip=\baselineskip


   \newbox\squarebox


   \newskip\sectionindent            \sectionindent=0pt
   \newskip\subsectionindent         \subsectionindent=0pt
  \def\thechapterhead{\relax}
  \def\thesectionhead{\relax}
  \def\thesubsecthead{\relax}
  \def\thesubsubsecthead{\relax}


   \def\GetIfExists #1 {
       \immediate\openin\testifexists=#1
       \ifeof\testifexists
           \immediate\closein\testifexists
       \else
         \immediate\closein\testifexists
         \input #1
       \fi
   }


   \def\stripbackslash#1#2*{\def\strippedname{#2}}

   \def\ifundefined#1{\expandafter\ifx\csname#1\endcsname\relax}

   \def\val#1{%
      \expandafter\stripbackslash\string#1*%
      \ifundefined{\strippedname}%
      \message{Warning! The control sequence \noexpand#1 is not defined.} ? %
      \else\csname\strippedname\endcsname\fi%
   }
%
%
   \def\CheckForOverWrite#1{%
      \expandafter\stripbackslash\string#1*%
      \ifundefined{\strippedname}%
      \else%
         \ifwarncsname
            \message{Warning! The control sequence \noexpand#1 is being
          overwritten.}%
          \else
          \fi
      \fi%
   }

   \def\FootNoteFonts{\Tenpoint}

   \def\Vfootnote#1{%
      \insert\footins%
      \bgroup%
         \interlinepenalty=\interfootnotelinepenalty%
         \floatingpenalty=20000%
         \singl@true\doubl@false%
         \FootNoteFonts%
         \splittopskip=\ht\strutbox%
         \boxmaxdepth=\dp\strutbox%
         \leftskip=\footindent%
         \rightskip=\z@skip%
         \parindent=0.5%
         \footindent%
         \parfillskip=0pt plus 1fil%
         \spaceskip=\z@skip%
         \xspaceskip=\z@skip%
         \footnotespecial%
         \Textindent{#1}%
         \footstrut%
         \futurelet\next\fo@t%
   }

   \def\csnamech@ck{%
       \ifcsnameopen%
       \else%
           \global\csnameopentrue%
           \immediate\openout\csnamewrite=\the\jobdir\jobname.csnames%
           \immediate\write\csnamewrite{\unlock}%
       \fi%
   }

   \def\linksch@ck{%
          \iflinkopen%
          \else%
              \global\linkopentrue%
              \immediate\openout\linkwrite=\the\jobdir\jobname.links%
          \fi%
   }

   \def\c@ntentscheck{%
       \ifcontentsopen%
       \else%
           \global\contentsopentrue%
           \immediate\openout\contentswrite=\the\jobdir\jobname.contents%
           \immediate\write\contentswrite{%
                \noexpand\titlestyle{Table of Contents}%
           }%
           \immediate\write\contentswrite{\noexpand\bigskip}%
       \fi%
   }

   \def\t@bleofplatescheck{%
       \ifplatesopen%
       \else%
           \global\platesopentrue%
           \immediate\openout\plateswrite=\the\jobdir\jobname.plates%
           \immediate\write\plateswrite{%
                \noexpand\titlestyle{Illustrations}%
           }%
           \immediate\write\plateswrite{%
              \unlock%
           }%
           \immediate\write\plateswrite{\noexpand\bigskip}%
       \fi%
   }

   \def\t@bleoftablescheck{%
       \iftableoftablesopen%
       \else%
           \global\tableoftablesopentrue%
          \immediate\openout\tableoftableswrite=\the\jobdir\jobname.tables%
           \immediate\write\tableoftableswrite{%
                \noexpand\titlestyle{Tables}%
           }%
           \immediate\write\tableoftableswrite{%
              \unlock%
           }%
           \immediate\write\tableoftableswrite{\noexpand\bigskip}%
       \fi%
   }


   \def\linkinput#1 {\input #1
       \iflinkedinput \relax \else \global\linkedinputtrue \fi
       \linksch@ck
       \immediate\write\linkwrite{#1}
   }


   \def\fil@#1 {%
       \ifdummypictures%
          \fileexistsfalse%
          \picfilename={}%
       \else%
          \fileexiststrue%
          \picfilename={#1}%
       \fi%
       \iflinkedinput%
          \iflinkopen \relax%
          \else%
            \global\linkopentrue%
            \immediate\openout\linkwrite=\the\jobdir\jobname.links%
          \fi%
          \immediate\write\linkwrite{#1}%
       \fi%
   }
   \def\Picture#1{%
      \gl@bal\advance\figurecount by 1%
      \CheckForOverWrite#1%
      \csnamech@ck%
      \immediate\write\csnamewrite{\def\noexpand#1{\the\figurecount}}%
      \xdef#1{\the\figurecount}\pl@t%
      \selfcaptionedtrue%
   }

   \def\s@vepicture{%
       \iffileexists \parsefilename \redopicturebox \fi%
       \ifdim\captionwidth>\z@ \else \captionwidth=\p@cwd \fi%
       \xdef\lastpicture{%
          \iffileexists%
             \setbox0=\hbox{\raise\the\yshift \vbox{%
                \moveright\the\xshift\hbox{\picturedefinition}}%
             }%
          \else%
             \setbox0=\hbox{}%
          \fi
          \ht0=\the\p@cht \wd0=\the\p@cwd \dp0=\the\p@cdp%
          \vbox{\hsize=\the\captionwidth \line{\hss\box0 \hss }%
          \ifcaptioned%
             \vskip\the\captionskip \noexpand\Tenpoint%
             \ifselfcaptioned%
                Figure~\the\figurecount.\enspace%
             \fi%
             \the\Caption%
          \fi }%
       }%
       \iftableofplates%
          \ifplatesopen%
          \else%
             \t@bleofplatescheck%
          \fi%
          \ifselfcaptioned%
             \immediate\write\plateswrite{%
                \noexpand\platetext{%
                \noexpand\item{\rm \the\figurecount .}%
                \the\Caption}{\the\pageno}%
             }%
          \else%
             \immediate\write\plateswrite{%
                \noexpand\platetext{\the\Caption}{\the\pageno}%
             }%
          \fi%
       \fi%
       \endgroup%
   }

   \def\platesout{%
      \ifplatesopen
         \immediate\closeout\plateswrite%
         \global\platesopenfalse%
      \fi%
      \input \jobname.plates%
      \lock%
   }

   \def\platetext#1#2{%
       \hbox to \hsize{\vbox{\hsize=.9\hsize #1}\hfill#2}%
       \vskip \tableoftablesskip \vskip\parskip%
   }


   \def\acksection#1{\par
      \ifnum\the\lastpenalty=30000\else \penalty-100\smallskip \fi
      \noindent\undertext{#1}\enspace \vadjust{\penalty5000}}

   \def\ack{\acksection{Acknowledgements:}}


   \def\pres@tpicture{%
       \gl@bal\linesabove=\linesabove
       \s@vepicture
       \setbox\picturebox=\vbox{
       \kern \linesabove\baselineskip \kern 0.3\baselineskip
       \lastpicture \kern 0.3\baselineskip }%
       \dimen@=\p@cht \dimen@i=\dimen@
       \advance\dimen@i by \pagetotal
       \par \ifdim\dimen@i>\pagegoal \vfil\break \fi
       \dimen@ii=\hsize
       \advance\dimen@ii by -\pictureindent \advance\dimen@ii by -\p@cwd
       \setbox0=\vbox to\z@{\kern-\baselineskip \unvbox\picturebox \vss }
   }

   \def\subspaces@t#1:#2;{%
      \baselineskip = \normalbaselineskip%
      \multiply\baselineskip by #1 \divide\baselineskip by #2%
      \lineskip = \normallineskip%
      \multiply\lineskip by #1 \divide\lineskip by #2%
      \lineskiplimit = \normallineskiplimit%
      \multiply\lineskiplimit by #1 \divide\lineskiplimit by #2%
      \parskip = \normalparskip%
      \multiply\parskip by #1 \divide\parskip by #2%
      \abovedisplayskip = \normaldisplayskip%
      \multiply\abovedisplayskip by #1 \divide\abovedisplayskip by #2%
      \belowdisplayskip = \abovedisplayskip%
      \abovedisplayshortskip = \normaldispshortskip%
      \multiply\abovedisplayshortskip by #1%
        \divide\abovedisplayshortskip by #2%
      \belowdisplayshortskip = \abovedisplayshortskip%
      \advance\belowdisplayshortskip by \belowdisplayskip%
      \divide\belowdisplayshortskip by 2%
      \smallskipamount = \skipregister%
      \multiply\smallskipamount by #1 \divide\smallskipamount by #2%
      \medskipamount = \smallskipamount \multiply\medskipamount by 2%
      \bigskipamount = \smallskipamount \multiply\bigskipamount by 4%
   }


   \def\makename#1{
       \global\makenametrue
       \global\tempnametoks={#1}
   }

   \def\nomakename#1{\relax}


   \def\savename#1{%
      \CheckForOverWrite{#1}%
      \csnamech@ck%
      \immediate\write\csnamewrite{\def\the\tempnametoks{#1}}%
   }

   \def\FootNoteFonts{\Tenpoint}

   \def\Vfootnote#1{%
      \insert\footins%
      \bgroup%
         \interlinepenalty=\interfootnotelinepenalty%
         \floatingpenalty=20000%
         \singl@true\doubl@false%
         \FootNoteFonts%
         \splittopskip=\ht\strutbox%
         \boxmaxdepth=\dp\strutbox%
         \leftskip=\footindent%
         \rightskip=\z@skip%
         \parindent=0.5%
         \footindent%
         \parfillskip=0pt plus 1fil%
         \spaceskip=\z@skip%
         \xspaceskip=\z@skip%
         \footnotespecial%
         \Textindent{#1}%
         \footstrut%
         \futurelet\next\fo@t%
   }
%

   \def\eqname#1{%
      \CheckForOverWrite{#1}%
      \rel@x{\pr@tect%
      \csnamech@ck%
      \ifnum\equanumber<0%
          \xdef#1{{\noexpand\f@m0(\number-\equanumber)}}%
          \immediate\write\csnamewrite{%
            \def\noexpand#1{\noexpand\f@m0 (\number-\equanumber)}}%
          \gl@bal\advance\equanumber by -1%
      \else%
          \gl@bal\advance\equanumber by 1%
          \ifusechapterlabel%
            \xdef#1{{\noexpand\f@m0(\ifcn@@ \chapterlabel.\fi%
               \number\equanumber)}%
            }%
          \else%
             \xdef#1{{\noexpand\f@m0(\ifcn@@%
                 {\the\chapterstyle{\the\chapternumber}}.\fi%
                 \number\equanumber)}}%
          \fi%
          \ifcn@@%
             \ifusechapterlabel
                \immediate\write\csnamewrite{\def\noexpand#1{(%
                  {\chapterlabel}.%
                  \number\equanumber)}%
                }%
             \else
                \immediate\write\csnamewrite{\def\noexpand#1{(%
                  {\the\chapterstyle{\the\chapternumber}}.%
                  \number\equanumber)}%
                }%
             \fi%
          \else%
              \immediate\write\csnamewrite{\def\noexpand#1{(%
                  \number\equanumber)}}%
          \fi%
      \fi}%
      #1%
   }

   \def\eqn{\eqno\eqname}

   \let\eqnalign=\eqname


   \def\REFNUM#1{%
      \CheckForOverWrite{#1} %
      \rel@x\gl@bal\advance\referencecount by 1%
      \xdef#1{\the\referencecount}%
      \csnamech@ck%
      \immediate\write\csnamewrite{\def\noexpand#1{\the\referencecount}}%
   }

   %

   \def\FIGNUM#1{
      \CheckForOverWrite{#1}%
      \rel@x\gl@bal\advance\figurecount by 1%
      \xdef#1{\the\figurecount}%
      \csnamech@ck%
      \immediate\write\csnamewrite{\def\noexpand#1{\the\figurecount}}%
   }


   \def\TABNUM#1{%
      \CheckForOverWrite{#1}%
      \rel@x \gl@bal\advance\tablecount by 1%
      \xdef#1{\the\tablecount}%
      \csnamech@ck%
      \immediate\write\csnamewrite{\def\noexpand#1{\the\tablecount}}%
   }


   \def\tableoftableson{%
      \global\tableoftablestrue%

      \gdef\TABLE##1##2{%
         \t@bleoftablescheck%
         \TABNUM ##1%
         \immediate\write\tableoftableswrite{%
            \noexpand\tableoftablestext{%
            \noexpand\item{\rm \the\tablecount .}%
                ##2}{\the\pageno}%
             }%
      }

      \gdef\Table##1{\TABLE\?{##1}Table~\?}
   }

   \def\tableoftablestext#1#2{%
       \hbox to \hsize{\vbox{\hsize=.9\hsize #1}\hfill#2}%
       \vskip \tableoftablesskip%
   }

   \def\tableoftablesout{%
      \iftableoftablesopen
         \immediate\closeout\tableoftableswrite%
         \global\tableoftablesopenfalse%
      \fi%
      \input \jobname.tables%
      \lock%
   }

%
%
%
%
%
%

   \def\contentsoff{\contentsonfalse}

   \def\f@m#1{\f@ntkey=#1\fam=\f@ntkey\the\textfont\f@ntkey\rel@x}
   \def\em@{\rel@x%
      \ifnum\f@ntkey=0\it%
      \else%
         \ifnum\f@ntkey=\bffam\it%
         \else\rm  %
         \fi%
      \fi%
   }

   \def\fontsoff{%
      \def\mit{\relax}%
      \let\oldstyle=\mit%
      \def\cal{\relax}%
      \def\it{\relax}%
      \def\sl{\relax}%
      \def\bf{\relax}%
      \def\tt{\relax}%
      \def\caps{\relax}%
      \let\cp=\caps%
   }


   \def\fontson{%
      \def\rm{\n@expand\f@m0}%
      \def\mit{\n@expand\f@m1}%
      \let\oldstyle=\mit%
      \def\cal{\n@expand\f@m2}%
      \def\it{\n@expand\f@m\itfam}%
      \def\sl{\n@expand\f@m\slfam}%
      \def\bf{\n@expand\f@m\bffam}%
      \def\tt{\n@expand\f@m\ttfam}%
      \def\caps{\n@expand\f@m\cpfam}%
      \let\cp=\caps%
   }

   \fontson
%


   \def\@alpha#1{\count255='140 \advance\count255 by #1\char\count255}
   \def\alphabetic{\@alpha}
   \def\@Alpha#1{\count255='100 \advance\count255 by #1\char\count255}
   \def\Alphabetic{\@Alpha}
   \def\@Roman#1{\uppercase\expandafter{\romannumeral #1}}
   \def\Roman{\@Roman}
   \def\@roman#1{\romannumeral #1}
   \def\roman{\@roman}
   \def\@number#1{\number #1}
   \def\Number{\@number}

   \def\leaderfill{\leaders\hbox to 1em{\hss.\hss}\hfill}

   \def\chapterinfo#1{%
      \line{%
         \ifcn@@%
            \hbox to \itemsize{\hfil\chapterlabel .\quad\ }%
         \fi%
         \noexpand{#1}\leaderfill\the\pagenumber%
      }%
   }

   \def\sectioninfo#1{%
      \line{%
         \ifcn@@%
            \hbox to 2\itemsize{\hfil\sectlabel \quad}%
          \else%
            \hbox to \itemsize{\hfil\quad}%
          \fi%
          \ \noexpand{#1}%
          \leaderfill \the\pagenumber%
      }%
   }

   \def\subsectioninfo#1{%
      \line{%
         \ifcn@@%
            \hbox to 3\itemsize{\hfil \quad\subsectlabel\quad}%
         \else%
            \hbox to 2\itemsize{\hfil\quad}%
         \fi%
          \ \noexpand{#1}%
          \leaderfill \the\pagenumber%
      }%
   }

   \def\subsubsecinfo#1{%
      \line{%
         \ifcn@@%
            \hbox to 4\itemsize{\hfil\subsubsectlabel\quad}%
         \else%
            \hbox to 3\itemsize{\hfil\quad}%
         \fi%
         \ \noexpand{#1}\leaderfill \the\pagenumber%
      }%
   }

   \def\CONTENTS#1;#2{
       {\let\makename=\nomakename
        \if#1C
            \immediate\write\contentswrite{\chapterinfo{#2}}%
        \else\if#1S
                \immediate\write\contentswrite{\sectioninfo{#2}}%
             \else\if#1s
                     \immediate\write\contentswrite{\subsectioninfo{#2}}%
                  \else\if#1x
                          \immediate\write\contentswrite{%
                              \subsubsecinfo{#2}}%
                       \fi
                  \fi
             \fi
        \fi
       }
   }

   \def\chapterreset{\gl@bal\advance\chapternumber by 1%
       \ifnum\equanumber<0 \else\gl@bal\equanumber=0 \fi%
       \gl@bal\sectionnumber=0  \gl@bal\let\sectlabel=\rel@x%
       \gl@bal\subsectnumber=0   \gl@bal\let\subsectlabel=\rel@x%
       \gl@bal\subsubsectnumber=0 \gl@bal\let\subsubsectlabel=\rel@x%
       \ifcn@%
           \gl@bal\cn@@true {\pr@tect\xdef\chapterlabel{%
           {\the\chapterstyle{\the\chapternumber}}}}%
       \else%
           \gl@bal\cn@@false \gdef\chapterlabel{\rel@x}%
       \fi%
       \gl@bal\startofchaptertrue%
   }

   \def\chapter#1{\par \penalty-300 \vskip\chapterskip%
       \spacecheck\chapterminspace%
       \gdef\thechapterhead{#1}%
       \gdef\thesectionhead{\relax}%
       \gdef\thesubsecthead{\relax}%
       \gdef\thesubsubsecthead{\relax}%
       \chapterreset \the\chapterformat{\the\chapterfontstyle%
          \ifcn@@\chapterlabel.~~\fi #1}%
       \nobreak\vskip\headskip \penalty 30000%
       {\pr@tect\wlog{\string\chapter\space \chapterlabel}}%
       \ifmakename%
           \csnamech@ck
           \ifcn@@%
              \immediate\write\csnamewrite{\def\the\tempnametoks{%
                 {\the\chapterstyle{\the\chapternumber}}}%
              }%
            \fi%
            \global\makenamefalse%
       \fi%
       \ifcontentson%
          \c@ntentscheck%
          \CONTENTS{C};{#1}%
       \fi%
       }%

   \def\section#1{\par \ifnum\lastpenalty=30000\else%
       \penalty-200\vskip\sectionskip \spacecheck\sectionminspace\fi%
       \gl@bal\advance\sectionnumber by 1%
       \gl@bal\subsectnumber=0%
       \gl@bal\let\subsectlabel=\rel@x%
       \gl@bal\subsubsectnumber=0%
       \gl@bal\let\subsubsectlabel=\rel@x%
       \gdef\thesectionhead{#1}%
       \gdef\thesubsecthead{\relax}%
       \gdef\thesubsubsecthead{\relax}%
       {\pr@tect\xdef\sectlabel{\ifcn@@%
          {\the\chapterstyle{\the\chapternumber}}.%
          {\the\sectionstyle{\the\sectionnumber}}\fi}%
       \wlog{\string\section\space \sectlabel}}%
       \the\sectionformat{\noindent\the\sectionfontstyle%
            {\ifcn@@\unskip\hskip\sectionindent\sectlabel~~\fi%
                \the\sectionfontstyleb#1}}%
       \par%
       \nobreak\vskip\headskip \penalty 30000%
       \ifmakename%
           \csnamech@ck%
           \ifcn@@%
              \immediate\write\csnamewrite{\def\the\tempnametoks{%
                 {\the\chapterstyle{\the\chapternumber}.%
                  \the\sectionstyle{\the\sectionnumber}}}
              }%
            \fi%
            \global\makenamefalse%
       \fi%
       \ifcontentson%
          \c@ntentscheck%
          \CONTENTS{S};{#1}%
       \fi%
   }

   \def\subsection#1{\par \ifnum\lastpenalty=30000\else%
       \penalty-200\vskip\sectionskip \spacecheck\sectionminspace\fi%
       \gl@bal\advance\subsectnumber by 1%
       \gl@bal\subsubsectnumber=0%
       \gl@bal\let\subsubsectlabel=\rel@x%
       \gdef\thesubsecthead{#1}%
       \gdef\thesubsubsecthead{\relax}%
       {\pr@tect\xdef\subsectlabel{\the\subsectionfontstyle%
           \ifcn@@{\the\chapterstyle{\the\chapternumber}}.%
           {\the\sectionstyle{\the\sectionnumber}}.%
           {\the\subsectstyle{\the\subsectnumber}}\fi}%
           \wlog{\string\section\space \subsectlabel}%
       }%
       \the\subsectionformat{\noindent\the\subsectionfontstyle%
         {\ifcn@@\unskip\hskip\subsectionindent%
          \subsectlabel~~\fi#1}}%
       \par%
       \nobreak\vskip\headskip \penalty 30000%
       \ifmakename%
           \csnamech@ck%
           \ifcn@@%
              \immediate\write\csnamewrite{\def\the\tempnametoks{%
                 {\the\chapterstyle{\the\chapternumber}}.%
                 {\the\sectionstyle{\the\sectionnumber}}.%
                 {\the\subsectstyle{\the\subsectnumber}}}%
              }%
            \fi%
            \global\makenamefalse%
       \fi%
       \ifcontentson%
          \c@ntentscheck%
          \CONTENTS{s};{#1}%
       \fi%
   }

   \def\subsubsection#1{\par \ifnum\lastpenalty=30000\else%
       \penalty-200\vskip\sectionskip \spacecheck\sectionminspace\fi%
       \gl@bal\advance\subsubsectnumber by 1%
       \gdef\thesubsubsecthead{#1}%
       {\pr@tect\xdef\subsubsectlabel{\the\subsubsectionfontstyle\ifcn@@%
           {\the\chapterstyle{\the\chapternumber}}.%
           {\the\sectionstyle{\the\sectionnumber}}.%
           {\the\subsectstyle{\the\subsectnumber}}.%
           {\the\subsubsectstyle{\the\subsubsectnumber}}\fi}%
           \wlog{\string\section\space \subsubsectlabel}%
       }%
       \the\subsubsectionformat{\the\subsubsectionfontstyle%
          \noindent{\ifcn@@\unskip\hskip\subsectionindent%
            \subsubsectlabel~~\fi#1}}%
       \par%
       \nobreak\vskip\headskip \penalty 30000%
       \ifmakename%
           \csnamech@ck%
           \ifcn@@%
              \immediate\write\csnamewrite{\def\the\tempnametoks{%
                {\the\chapterstyle{\the\chapternumber}.%
                 \the\sectionstyle{\the\sectionnumber}.%
                 \the\subsectionstyle{\the\subsectnumber}.%
                 \the\subsubsectstyle{\the\subsubsectnumber}}}%
              }%
            \fi%
            \global\makenamefalse%
       \fi%
       \ifcontentson%
          \c@ntentscheck%
          \CONTENTS{x};{#1}%
       \fi%
   }%

   \def\contentsinput{%
       \ifcontentson%
           \contentsopenfalse%
           \immediate\closeout\contentswrite%
           \global\oldheadline=\headline%
           \global\headline={\hfill}%
           \global\oldfootline=\footline%
           \global\footline={\hfill}%
           \fontsoff \unlock%
           \input \the\jobdir\jobname.contents%
           \fontson%
           \lock%
           \endpage%
           \global\headline=\oldheadline%
           \global\footline=\oldfootline%
       \else%
           \relax%
       \fi%
   }


       \def\phyzzxfootline{
           \footline={\ifletterstyle\the\letterfootline%
               \else\the\paperfootline\fi}%
       }

%

   {\obeyspaces}

   \def\verbfile#1{
       {\catcode`\\=12\catcode`\{=12
       \catcode`\}=12\catcode`\$=12\catcode`\&=12
       \catcode`\#=12\catcode`\%=12\catcode`\~=12
       \catcode`\_=12\catcode`\^=12\obeyspaces\obeylines\tt
       \verbdonetrue\openin\verbinfile=#1
       \loop\read\verbinfile to \inline
           \ifeof\verbinfile
               \verbdonefalse
           \else
              \leftline{\inline}
           \fi
       \ifverbdone\repeat
       \closein\verbinfile}
   }

   \def\boxit#1{\vbox{\hrule\hbox{\vrule\kern3pt%
       \vbox{\kern3pt#1\kern3pt}\kern3pt\vrule}\hrule}%
   }

   \def\square{%
      \setbox\squarebox=\boxit{\hbox{\phantom{x}}}
      \squareht = 1\ht\squarebox
      \squarewd = 1\wd\squarebox
      \vbox to 0pt{
          \offinterlineskip \kern -.9\squareht
          \hbox{\copy\squarebox \vrule width .2\squarewd height .8\squareht
              depth 0pt \hfill
          }
          \hbox{\kern .2\squarewd\vbox{%
            \hrule height .2\squarewd width \squarewd}
          }
          \vss
      }
   }

   \def\fboxit#1#2{
       \vbox{\hrule height #1
           \hbox{\vrule width #1
               \kern3pt \vbox{\kern3pt#2\kern3pt}\kern3pt \vrule width #1
           }
           \hrule height #1
       }
   }

   \let\eqnameold=\eqname

   \def\draft{\def\eqname##1{\eqnameold##1:{\tt\string##1}}
      \let\eqnalign = \eqname
   }
%
%
   \def\runningrightheadline{%
       \hfill%
       \tenit%
       \ifstartofchapter%
          \global\startofchapterfalse%
       \else%
          \ifcn@@ \the\chapternumber.\the\sectionnumber\quad\fi%
              {\fontsoff\thesectionhead}%
       \fi%
       \qquad\twelverm\folio%
   }

   \def\runningleftheadline{%
      \twelverm\folio\qquad%
      \tenit%
      \ifstartofchapter%
          \global\startofchapterfalse%
      \else%
         \ifcn@@%
             Chapter \the\chapternumber \quad%
         \fi%
         {\fontsoff\thechapterhead}%
         \hfill%
      \fi%
   }

   \runningheadlines={%
      \ifodd\pageno%
         \runningrightheadline%
      \else%
         \runningleftheadline%
      \fi
   }

%
%
%
%
%

   \font\dfont=cmr10 scaled \magstep5


   \newbox\cstrutbox
   \newbox\dlbox
   \newbox\vsk

   \setbox\cstrutbox=\hbox{\vrule height10.5pt depth3.5pt width\z@}

   \def\cstrut{\relax\ifmmode\copy\cstrutbox\else\unhcopy\cstrutbox\fi}

   \def\dl #1{\noindent\strut
       \setbox\dlbox=\hbox{\dfont #1\kern 2pt}%
       \setbox\vsk=\hbox{(}%
       \hangindent=1.1\wd\dlbox
       \hangafter=-2
       \strut\hbox to 0pt{\hss\vbox to 0pt{%
         \vskip-.75\ht\vsk\box\dlbox\vss}}%
       \noindent
   }

%
%

   \newdimen\fullhsize

   \fullhsize=6.5in
   \def\fullline{\hbox to\fullhsize}
   \let\l@r=L

   \newbox\leftcolumn
   \newbox\midcolumn

   \def\twocols{\hsize = 3.1in%
%
%
%
%
%
      \doublecolskip=.3333em plus .3333em minus .1em
      \global\spaceskip=\doublecolskip%
      \global\hyphenpenalty=0
      \singlespace
      \gdef\makeheadline{%
          \vbox to 0pt{ \skip@=\topskip%
          \advance\skip@ by -12pt \advance\skip@ by -2\normalbaselineskip%
          \vskip\skip@%
          \fullline{\vbox to 12pt{}\the\headline}\vss}\nointerlineskip%
      }%
      \def\makefootline{\baselineskip = 1.5\normalbaselineskip
           \fullline{\the\footline}
      }
      \output={%
          \if L\l@r%
             \global\setbox\leftcolumn=\columnbox \global\let\l@r=R%
          \else%
              \doubleformat \global\let\l@r=L%
          \fi%
          \ifnum\outputpenalty>-20000 \else\dosupereject\fi%
      }
      \def\doubleformat{
          \shipout\vbox{%
             \makeheadline%
             \fullline{\box\leftcolumn\hfil\columnbox}%
             \makefootline%
          }%
          \advancepageno%
      }
      \def\columnbox{\leftline{\pagebody}}
      \outer\def\twobye{%
          \par\vfill\supereject\if R\l@r \null\vfill\eject\fi\end%
      }%
   }

   \def\threecols{
       \hsize = 2.0in \tenpoint

      \doublecolskip=.3333em plus .3333em minus .1em
      \global\spaceskip=\doublecolskip%
      \global\hyphenpenalty=0

       \singlespace

       \def\makeheadline{\vbox to 0pt{ \skip@=\topskip
           \advance\skip@ by -12pt \advance\skip@ by -2\normalbaselineskip
           \vskip\skip@ \fullline{\vbox to 12pt{}\the\headline} \vss
           }\nointerlineskip
       }
       \def\makefootline{\baselineskip = 1.5\normalbaselineskip
                 \fullline{\the\footline}
       }

       \output={
          \if L\l@r
             \global\setbox\leftcolumn=\columnbox \global\let\l@r=M
          \else \if M\l@r
                   \global\setbox\midcolumn=\columnbox
                   \global\let\l@r=R
                \else \tripleformat \global\let\l@r=L
                \fi
          \fi
          \ifnum\outputpenalty>-20000 \else\dosupereject\fi
       }

       \def\tripleformat{
           \shipout\vbox{
               \makeheadline
               \fullline{\box\leftcolumn\hfil\box\midcolumn\hfil\columnbox}
               \makefootline
           }
           \advancepageno
       }

       \def\columnbox{\leftline{\pagebody}}

       \outer\def\threebye{
           \par\vfill\supereject
           \if R\l@r \null\vfill\eject\fi
           \end
       }
   }


%
%
%


   \everyjob{%

       \immediate\openin\testifexists=m
       \ifeof\testifexists
           \immediate\closein\testifexists
       \else
         \immediate\closein\testifexists
         \input m
       \fi
   yphyx.tex
      \ifforwardrefson%
         
       \immediate\openin\testifexists=\the
       \ifeof\testifexists
           \immediate\closein\testifexists
       \else
         \immediate\closein\testifexists
         \input \the
       \fi
   \jobdir\jobname.csnames
      \fi%
   }

\contentsoff
\lock
%
%
%
\unlock

   \newtoks\blank
   \newtoks\dither
   \newtoks\resolution
   \newtoks\smooth
   \newtoks\nonlinear
   \newtoks\threshold
   \newtoks\amigapicheight
   \newtoks\amigapicwidth
   \newtoks\amigapicdepth
   \newdimen\scaledheight
   \newdimen\scaledwidth
   \newdimen\scaleddepth
   \newread\epsffilein    
   \newif\ifepsffileok    
   \newif\ifepsfbbfound   
   \newif\ifepsfverbose   
   \newif\ififftoimp      
   \newdimen\epsfxsize    
   \newdimen\epsfysize    
   \newdimen\epsftsize    
   \newdimen\epsfrsize    
   \newdimen\epsftmp      
   \newdimen\pspoints     

   \pspoints=1bp          
   \epsfxsize=0pt         
   \epsfysize=0pt         
   \def\epsfbox#1{%
      \global\def\epsfllx{72}%
      \global\def\epsflly{72}%
      \global\def\epsfurx{540}%
      \global\def\epsfury{720}%
      \def\lbracket{[}%
      \def\testit{#1}%
      \ifx\testit\lbracket%
         \let\next=\epsfgetlitbb
      \else%
         \let\next=\epsfnormal%
      \fi%
      \next{#1}
   }%
   \def\epsfgetlitbb#1#2 #3 #4 #5]#6{%
      \epsfgrab #2 #3 #4 #5 .\\%
      \epsfsetgraph{#6}%
   }
   \def\epsfnormal#1{%
      \epsfgetbb{#1}%
      \epsfsetgraph{#1}%
   }
%
%
%
   \def\epsfgetbb#1{%
      \openin\epsffilein=#1 %
      \ifeof\epsffilein%
         \errmessage{I couldn't open #1 , will ignore it}%
      \else%
         {%
            \epsffileoktrue%
            \chardef\other=12%
            \def\do##1{\catcode`##1=\other}%
            \dospecials%
            \catcode`\ =10%
            \loop%
               \read\epsffilein to \epsffileline
               \ifeof\epsffilein%
                  \epsffileokfalse%
               \else%
            %
            %
                  \expandafter\epsfaux\epsffileline:. \\%
               \fi%
            \ifepsffileok\repeat%
            \ifepsfbbfound%
            \else%
               \ifepsfverbose%
                  \message{No bounding box comment in #1; using defaults}%
               \fi%
            \fi%
         }%
         \closein\epsffilein%
      \fi%
   }%
%
%
   \def\epsfsetgraph#1{%
      \epsfrsize=\epsfury\pspoints%
      \advance\epsfrsize by-\epsflly\pspoints%
      \epsftsize=\epsfurx\pspoints%
      \advance\epsftsize by-\epsfllx\pspoints
       %
       %
      \epsfxsize\epsfsize\epsftsize\epsfrsize
      \ifnum\epsfxsize=0%
         \ifnum\epsfysize=0%
            \epsfxsize=\epsftsize%
            \epsfysize=\epsfrsize%
            %
            %
         \else%
            \epsftmp=\epsftsize%
            \divide\epsftmp\epsfrsize%
            \epsfxsize=\epsfysize%
            \multiply\epsfxsize\epsftmp%
            \multiply\epsftmp\epsfrsize%
            \advance\epsftsize-\epsftmp%
            \epsftmp=\epsfysize%
            \loop%
               \advance\epsftsize\epsftsize%
               \divide\epsftmp 2%
            \ifnum\epsftmp>0
               \ifnum\epsftsize<\epsfrsize%
               \else%
                  \advance\epsftsize-\epsfrsize%
                  \advance\epsfxsize\epsftmp%
               \fi%
            \repeat%
         \fi%
      \else%
         \epsftmp=\epsfrsize%
         \divide\epsftmp\epsftsize%
         \epsfysize=\epsfxsize%
         \multiply\epsfysize\epsftmp%
         \multiply\epsftmp\epsftsize%
         \advance\epsfrsize-\epsftmp%
         \epsftmp=\epsfxsize%
         \loop%
            \advance\epsfrsize\epsfrsize%
            \divide\epsftmp 2%
         \ifnum\epsftmp>0
            \ifnum\epsfrsize<\epsftsize%
            \else%
               \advance\epsfrsize-\epsftsize%
               \advance\epsfysize\epsftmp%
            \fi
         \repeat%
      \fi%
      %
      %
      \ifepsfverbose%
         \message{#1: width=\the\epsfxsize, height=\the\epsfysize}%
      \fi
      \epsftmp=10\epsfxsize%
      \divide\epsftmp\pspoints%
   }%
%
%
{\catcode`\%=12 \global\let\epsfpercent=
   %
   %
   \long\def\epsfaux#1#2:#3\\{%
      \ifx#1\epsfpercent
         \def\testit{#2}%
         \ifx\testit\epsfbblit%
            \epsfgrab #3 . . . \\%
            \epsffileokfalse%
            \global\epsfbbfoundtrue%
         \fi%
      \else%
         \ifx#1%
            \par%
         \else%
            \epsffileokfalse%
         \fi%
      \fi%
   }%
%
%
   \def\epsfgrab #1 #2 #3 #4 #5\\{%
      \global\def\epsfllx{#1}%
      \ifx\epsfllx\empty%
         \epsfgrab #2 #3 #4 #5 .\\%
      \else
         \global\def\epsflly{#2}%
         \global\def\epsfurx{#3}%
         \global\def\epsfury{#4}%
      \fi%
   }%
%
%
   \def\epsfsize#1#2{\epsfxsize}%
%
%
   %

   %

   \def\parsefilename{%
      \ifreadyfile%
      \else%
         \iffigscaleopen%
         \else%
            \gl@bal\figscaleopentrue
            \immediate\openout\figscalewrite=\jobname.scalecon%
         \fi
         \toks0={ }%
         \immediate\write\figscalewrite{%
            \the\p@cwd \the\toks0 \the\p@cht \the\toks0 \the\picfilename}%
         \expandafter\p@rse \the\picfilename..\endp@rse%
       \fi%
    }%

   \def\p@rse#1.#2.#3\endp@rse{\relax}

   \def\dop@rse#1.#2..{\relax}

   \def\cutdimen{\scaledheight=1\ht\picturebox
      \scaledwidth=1\wd\picturebox
      \scaleddepth=1\dp\picturebox
      \expandafter\myp@rse\the\scaledheight[\noexpand\amigapicheight]\endp@rse
      \expandafter\myp@rse\the\scaledwidth[\noexpand\amigapicwidth]\endp@rse
      \expandafter\myp@rse\the\scaleddepth[\noexpand\amigapicdepth]\endp@rse
   }

   \def\myp@rse#1.#2[#3]\endp@rse{#3 {#1}}

   \blank={ }
   \smooth={0}
   \dither={classic}
   \resolution={3x3}
   \nonlinear={(0,0),(1,1)}
   \threshold={.8}

   \def\redopicturebox{%
      \ififftoimp%
         \cutdimen%
      \else%
         \ifgrayscale%
            \cutdimen%
         \else%
            \epsfxsize=\the\width%
            \epsfbox{\the\picfilename}%
            \global\height=\epsfysize
            \global\width=\epsfxsize
         \fi%
      \fi%
      \edef\picturedefinition{%
         \ifgrayscale%
            \special{%
               ifffile=\the\picfilename \the\blank%
               hscale=\the\amigapicwidth 00 \the\blank%
               vscale=\the\amigapicheight 00 \the\blank%
               smooth=\the\smooth \the\blank%
               \the\dither =\the\resolution  \the\blank%
               nonlinear=\the\nonlinear%
            }%
         \else%
            \ififftoimp%
               \special{%
                  insert(\the\picfilename)%
               }%
            \else
               \special{%
                  PSfile=\the\picfilename \the\blank%
                  llx=\epsfllx\the\blank lly=\epsflly\the\blank%
                  urx=\epsfurx\the\blank ury=\epsfury\the\blank %
                  rwi=\number\epsftmp%
               }%
            \fi%
         \fi%
      }%
      \global\epsfxsize=0pt%
      \global\epsfysize=0pt%
   }%

\lock
\def\address#1{\par\kern 4.16667pt\noindent%
\titlestyle{\twelvepoint\it #1}}
\def\SLACADD{\address{Stanford Linear Accelerator Center\break
      Stanford University, Stanford, California, 94309}}

\def\pigg{\Pi_{\gamma\gamma}}
\def\piggp{\Pi_{\gamma\gamma}^\prime}
\def\dalpha{\Delta\alpha}
\def\dalphamz{\Delta\alpha(M_Z^2)}
\def\dalhad{\Delta\alpha_{had}}
\def\dalhadmz{\Delta\alpha_{had}(M_Z^2)}
\def\dalexp{\delta(\Delta\alpha_{had})_{exp}}
\def\dalparam{\delta(\Delta\alpha_{had})_{param}}
\def\rhad{R_{had}}
\def\rtot{R_{tot}}
\def\rfit{R_{fit}}
\def\rqcd{R_{QCD}}
\def\rhadt{R_{had}^{\ge3}}
\def\ee{e^+e^-}
\def\roots{\sqrt{s}}

\def\swein{\sin^2\theta_W^{\rm eff}}
\def\etal{\it et al.\rm}
\def\z0{Z}

\def\oversim#1#2{\lower0.7ex\vbox{\baselineskip 0pt plus0pt minus0pt
\lineskip 0pt plus0pt minus0pt \lineskiplimit 0pt
  \ialign{$\mathsurround=0pt #1\hfil##\hfil$\crcr#2\crcr\sim\crcr}}}
\physrev
\tolerance=10000
 \normalspace
\pubnum{95-7001}
\date{November 1995}
\pubtype={(T/E)}
\titlepage
 \title{{\fourteenbf
 Reevaluation of the Hadronic Contribution to $\alpha(M_Z^2)$}$^*$}
\foot{Updated version of SLAC-PUB-6710.  Work supported by Department
of Energy, contract DE-AC03-76SF00515.}
\author{Morris L. Swartz}
\SLACADD
\vskip 0.2in
 %
 %
 \abstract
We reevaluate the hadronic part of the electromagnetic vacuum
expectation value using the standard dispersion integral approach that
utilizes the hadronic cross section measured in $\ee$ experiments as
input.  Previous analyses are based upon point-by-point trapezoidal
integration which does not treat experimental errors in an optimal way.
We use a technique that weights the experimental inputs by their stated
uncertainties, includes correlations, and incorporates some
refinements.  We find the five-flavor hadronic contribution to the
fractional change in the electromagnetic coupling constant at
$q^2=M_Z^2$, $\dalphamz$, to be $0.02752\pm0.00046$,
which leads to a value of the electromagnetic coupling constant,
$\alpha^{-1}(M_Z^2) = 128.96\pm0.06$.

 \submit{Physical Review D}
 \endpage

\sequentialequations
\normalspace
\chapter{Introduction}

At the current time, a large program of precise electroweak
measurements is being conducted throughout the world.  The object of
this program is to test the electroweak Standard Model by comparing the
measured values of a large set of electroweak observables with the
predictions of the Minimal Standard Model (MSM).  The Standard Model
calculations have been performed to full one-loop accuracy and partial
two-loop precision by a large community of researchers.  In all of
these calculations, it is necessary to evaluate the
one-particle-irreducible contributions to the photon self-energy
$\pigg(q^2)$ or the related quantity
$\piggp(q^2)\equiv(\pigg(q^2)-\pigg(0))/q^2$ at the $Z$ mass scale
$q^2=M_Z^2$.  These quantities are usually absorbed into the definition
of the running electromagnetic coupling $\alpha(q^2)$,
$$\alpha(q^2)\equiv{\alpha_0\over1-\left[\piggp(q^2)-\piggp(0)\right]},
\eqn\alrundef    $$
where $\alpha_0=1/137.0359895(61)$ is the electromagnetic fine
structure constant.  This quantity is also represented as the
fractional change in the electromagnetic coupling constant $\dalpha$,
$$\dalpha(q^2) =
{\alpha(q^2)-\alpha_0\over\alpha(q^2)}=\piggp(q^2)-\piggp(0).
\eqn\daldef $$

Using analytic techniques and the optical theorem applied to the
amplitude for s-channel Bhabha scattering, the quantity $\dalpha$ has
been related to the cross section for the process $\ee\to\gamma^*\to
{\rm all}$ ($\sigma_{tot}$) as follows%
\ref{N.~Cabibbo and R.~Gatto, {\it Phys. Rev.} {\bf 124}, 1577
(1961).},
$$\dalpha(q^2)={\alpha_0\over3\pi}{\rm P}\int_{4m_e^2}^\infty
ds{q^2\over s(q^2-s)}\rtot(s), \eqn\dispdef $$
where $\rtot(s)$ is the ratio of the total cross section to the
(massless) muon pair cross section
$\sigma_{\mu\mu}(s)=4\pi\alpha^2(s)/3s$
at the center-of-mass energy $\sqrt{s}$.  The cross section
$\sigma_{tot}$ is the physical cross section which has been corrected
for initial state radiation.  The actual quantity measured in most
experiments is discussed in Appendix A.  It should be noted in passing
that equation~\dispdef\ is correct to all orders in $\alpha_0$ and
relies only upon the assumption that the real part of $\pigg$ is much
larger than its imaginary part (the next-order correction is
proportional to ${\rm Im}^2\pigg/|\pigg|^2$ which is approximately
3$\times$10$^{-4}$ at $q^2=M_Z^2$).  It is straightforward to evaluate
equation~\dispdef\ for the continuum leptonic cross sections%
\ref{See G.~Burgers and W.~Hollik, CERN-TH-5131/88, August 1988, and
CERN 88-06, September 1988.}.
In the limit that the scale $q^2$ is much larger than the square of the
lepton mass $m_\ell^2$, the contribution of the continuum leptonic
cross sections is given by the following expression,
$$\dalpha_{\ell}(q^2)={\alpha_0\over3\pi}\sum_\ell\left[-{5\over3}+{\rm
ln} {q^2\over m_\ell^2}\right].  \eqn\dallep   $$

The remaining contributions to $\rtot$ consist of the continuum
hadronic cross section and the $J^P=1^-$ resonances and are labelled
$\rhad$.  Since the cross sections for the resonances and low energy
continuum are not accurately calculable from first principles,
experimental inputs are used to evaluate their contributions
equation~\dispdef.  The contribution of open top quark production to
the integral is accurately calculable and since the top quark mass is
not known precisely, only the five flavor hadronic cross section is
included in $\rhad$.  The corresponding contribution to $\dalpha(q^2)$
is therefore,
$$\dalhad(q^2)={\alpha_0\over3\pi}{\rm P}\int_{4m_\pi^2}^\infty
ds{q^2\over s(q^2-s)}\rhad(s). \eqn\dispint $$

\REF\old{F.A.~Berends and G.J.~Komen, {\it Phys. Lett.} {\bf 63B}, 432
(1976); F.~Jegerlehner, {\it Z. Phys.} {\bf C32}, 195 and 425 (1986);
B.W.~Lynn, G.~Penso, and C.~Verzegnassi, {\it Phys. Rev.} {\bf D35}, 42
(1987).}
\REF\burk{H.~Burkhardt, F.~Jegerlehner, G.~Penso, and C.~Verzegnassi,
{\it Z. Phys.} {\bf C43}, 497 (1989).}
\REF\jeger{F.~Jegerlehner, {\it Progress in Particle and Nuclear
Physics}, Vol~27, ed. A.~Faessler, Pergamon Press, Oxford 1991, p.~32;
F.~Jegerlehner, {\it Proceedings of the Theoretical Advanced Study
Institute in Elementary Particle Physics}, Boulder, 1990, ed. M.~Cvetic
and P.~Langacker, World Scientific, Teaneck N.J. 1991, p.~476.}
\REF\mz{A.D.~Martin and D.~Zeppenfeld, {\it Phys. Lett.} {\bf B345},
558 (1995).}
\REF\ej{S.~Eidelman and F.~Jegerlehner, PSI-PR-95-1, BudkerINP 95-5,
January 1995.}
\REF\bp{H.~Burkhardt and B.~Pietrzyk, PSI-PR-95-1, LAPP-EXP-95.05, June
1995.}
Equation~\dispint\ has been evaluated at the $Z$ boson mass scale a
number of times\refmark{\old-\bp}.
 The most recent evaluations are by Martin and Zeppenfeld\refmark{\mz},
Eidelman and Jegerlehner\refmark{\ej}, and by Burkhardt and
Pietrzyk\refmark{\bp} yield
$$\dalhadmz=\cases{0.02739\pm0.00042,&Reference~\mz\cr
0.0280\pm0.0007,&Reference~\ej\cr
0.0280\pm0.0007,&Reference~\bp.\cr} \eqn\dalcur       $$
The authors of Reference~\mz\ use perturbative QCD to parameterize the
continuum $\rhad(s)$ above $\sqrt{s}=3$~GeV and linear interpolation of
measured data below that point.  The two-body final states $\pi^+\pi^-$
and $K^+K^-$ are fit to parameterizations which include the $\rho$,
$\omega$, and $\phi$ resonances.  The remaining resonance contributions
are calculated from an analytic expression which results from
integrating a Breit-Wigner lineshape and depends upon the masses,
widths, and leptonic widths of each resonance.  The authors of
Reference~\ej\ use linear interpolation (trapezoidal integration) of
measured data points to evaluate the continuum, $\pi^+\pi^-$, and
$K^+K^-$ contributions.  Above $\sqrt{s}=40$~GeV, they use perturbative
QCD to evaluate $\rhad$.  The contributions of the $\omega$, $\phi$,
$J/\psi$-family, and $\Upsilon$-family resonances are included by
integrating a Breit-Wigner lineshape.  The authors of Reference~\bp\
use smoothed averages of data to evaluate the continuum contribution, a
parameterization to evaluate the $\pi^+\pi^-$ contribution, and the
analytic expression to evaluate the contribution of the remaining
resonances.

This document reports on an evaluation of equation~\dispint\ which is
performed in a somewhat different way from those listed above.  In
particular, the technique employed makes better use of the information
provided by the various $\rhad$ measurements, avoids some pitfalls
inherent in the trapezoidal technique, and naturally provides an
accurate estimate of the uncertainty on the result.  We find
$$\dalhadmz = 0.02752\pm0.00046,       $$
which appears to be consistent with Refs.~\mz-\bp\ within quoted
errors.

The result reported here updates an earlier result%
\Ref\mls{M.L.~Swartz, SLAC-PUB-6710, December 1994, hep-ph 9411353.}
which was more discrepant with Refs.~\mz-\bp.  The updated value of
$\dalhadmz$ is larger than the previous one by $8.6\times10^{-4}$ for
five reasons.  The previous analysis used the six-flavor definition of
$\dalhad$ which differs from the five-flavor quantity by
$0.6\times10^{-4}$.  A (hopefully) less controversial choice of
$\alpha_s(M_Z^2)$ shifts the result by $-0.5\times10^{-4}$.  The
fitting procedure used in the previous analysis was biased toward
smaller $\rhad$ values; correction of this problem gives a difference
of $2.9\times10^{-4}$.  Small corrections to the analysis of the
resonant contribution change the result by $-0.1\times10^{-4}$.  But,
the largest change is caused by the incorporation of a precise, new
measurement of $\rhad$ near charm threshold which alters the result by
$5.8\times10^{-4}$.  Although the net result is somewhat closer to
those given above, a detailed comparison of the actual integrated cross
section with one used in a trapezoidal integration (see Section~2.7)
indicates that significant differences persist.

\chapter{The Analysis}

Any attempt to combine the results of many experiments is a perilous
undertaking.  Many different techniques and approaches have been used.
Not all researchers have addressed all possible problems nor are
systematic error estimates performed in uniform ways or to uniform
standards.  We therefore adopt some the techniques of the
Particle Data Group%
\Ref\pdg{Review of Particle Properties: L.~Montanet, \etal, {\it Phys.
Rev.} {\bf D50}, 1173 (1994).}.
Older measurements which are contradicted by newer, more precise work
are excluded from the analysis.  Parameter uncertainties that are
extracted from fits with $\chi^2$ per degree of freedom (dof) larger
than one are rescaled by the factor $\sqrt{\chi^2/dof}$.

\section{Analysis Technique}

The experimental measurements of $\rhad(s)$ are performed over limited
regions of $W\equiv\roots$.  Typically, an experimental result consists
of several points $\rhad^i=\rhad(W_i)$ measured at closely spaced
energy points $W_i$.  Each set of measurements is accompanied by a set
of point-to-point uncertainties (statistical and systematic)
$\sigma_i({\rm ptp})$ and an overall normalization uncertainty
$\sigma({\rm norm})$.  Quite often, the point-to-point uncertainties
are much smaller than the normalization uncertainty.  {\it A typical
experimental result therefore consists of an accurately measured shape
of less certain normalization}.  In this case, the values of the
measured points are strongly intercorrelated.  For future reference, we
label these as Type~I correlations.

The normalization uncertainties usually incorporate purely
detector-related effects, acceptance uncertainties, and uncertainties
on radiative corrections and background corrections.  The largest
normalization uncertainties (15-20\%) are associated with the oldest
measurements of $\rhad$ in the $W=1-5$~GeV region.  These experiments
typically had limited acceptance which when combined with a (common)
limited understanding of the event structure lead to large
uncertainties in the overall detection efficiencies.  {\it The
normalization errors associated with different sets of measurements
performed at similar energies and times may be strongly correlated.}
These correlations are distinct from those discussed above (which must
be present) and are labelled as Type~II correlations.  When combining
the results of separate experiments, one must be careful to include the
possible presence of Type~II correlations in a conservative estimate of
the overall experimental uncertainty.

Most previous analyses of $\dalhad$ evaluate various contributions to
equation~\dispint\ by performing a trapezoidal integration with
measured values of $\rhad$.  Different data sets are combined by
weighting nearby points by the quadrature sums of their point-to-point
and normalization uncertainties (assuming that all points are
uncorrelated).  The effects of possible Type~II correlations on the
overall uncertainty are accounted for differently in different
analyses.  Eidelman and Jegerlehner\refmark{\ej} sum the uncertainties
associated with each point linearly.  Burkhardt and
Pietrzyk\refmark{\bp} and most of the earlier analyses assign typical
normalization uncertainties to various intervals in $W$ and sum the
corresponding uncertainties on $\dalhad$ in quadrature.  The use of
trapezoidal integration has two advantages: it is unbiased by human
prejudice about the functional form of $\rhad(s)$, and it would
automatically account for undiscovered resonances which are broad as
compared with the spacing of measurements.  Unfortunately, this
technique also has a serious shortcoming: it ignores the Type~I
correlations present in each data set.
\Picture\shape
\file={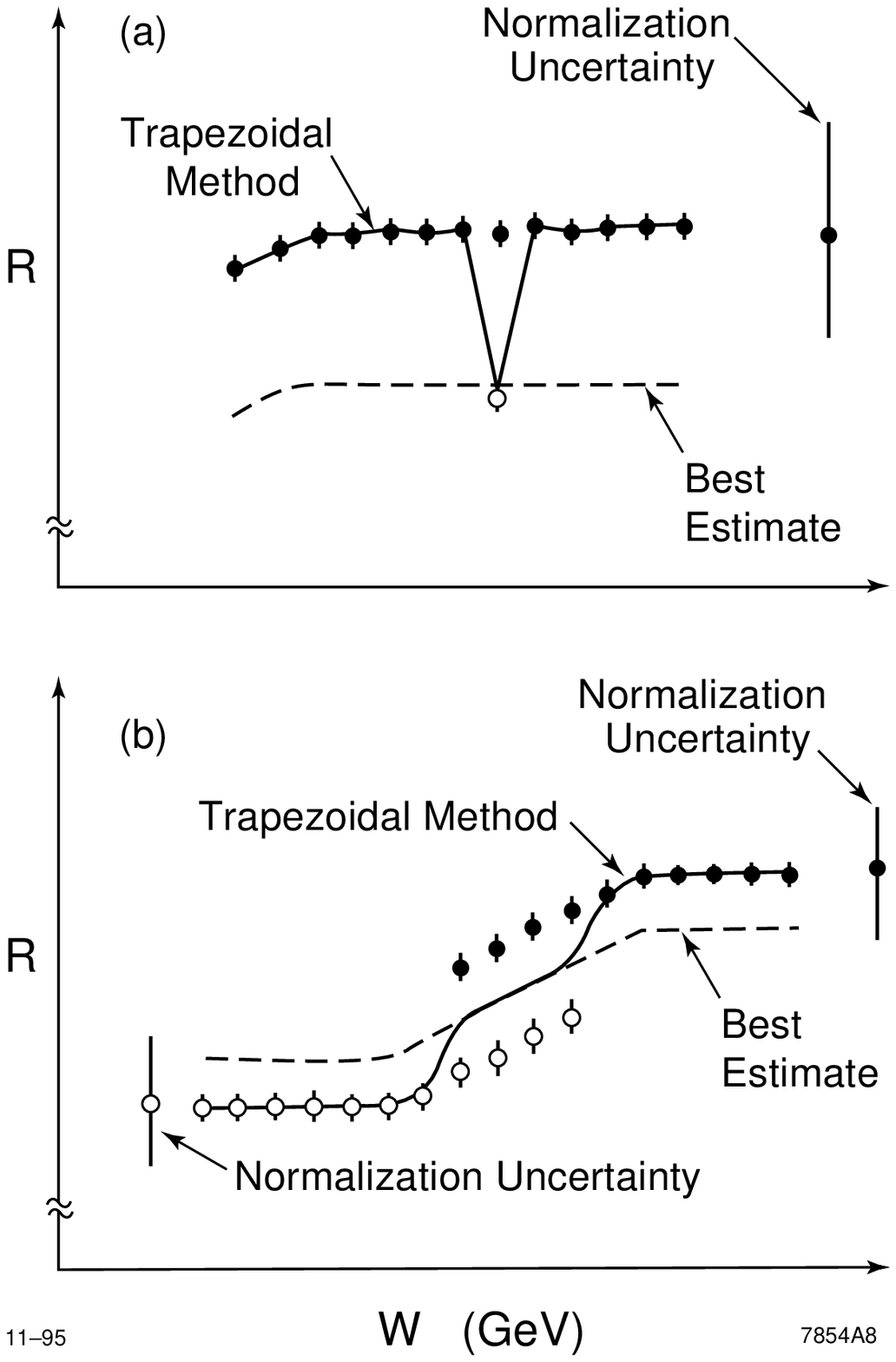}
\width=3.0in
\caption={
Two examples of the shape information loss inherent in the averaging
procedures used by trapezoidal analyses.}
\savepicture\pishape

Treating the combined (normalization and point-to-point)
uncertainties on the points in each set as independent loses the
(often precise) shape information associated with the set.  Two
examples of the loss of shape information are illustrated
Figure~\shape.  In part a), a data set with small point-to-point
errors (shown as solid dots) and a large normalization
uncertainty (illustrated to the right of the data) is combined
with a single precise measurement (shown as the open dot).  The
statistical averaging procedure used in the trapezoidal
integrations would yield the function shown as the solid curve.
The shape defined by the solid dots would be distorted near the
single precise point and the accurate normalization information
contained in the single measurement would be ignored.  An more
optimal procedure would use the shape information provided by the
solid dots and the normalization information provided by the open
dot yielding the dashed curve.
\vskip 0.15in
\centerline{\pishape}

Part b) of Figure~\shape\ shows
the result of combining two partly overlapping sets which have
small point-to-point uncertainties and large normalization
uncertainties (shown as open and solid dots, respectively).
In the region of overlap, the sets define a
consistent shape but differ in normalization.  An optimal
averaging procedure would average the normalizations and produce
the dashed curve.  The procedure adopted as part of the
trapezoidal analyses would yield the solid curve which agrees
with the dashed one only in the region of overlap and does not
preserve the shape determined by the data sets.  The trapezoidal
analyses described in References~\burk, \jeger, and \ej\ are
checked by first integrating individual data sets and then by
averaging the integrals.  While it might appear that this
procedure preserves shape information, the actual averaging of
the integrals can be carried out only in energy intervals where
the data sets overlap.  The net result therefore looks much like
the solid curve in part b).  It is not surprising that consistent
results were obtained.  Optimal use of the shape information can
occur only in techniques that allow the normalizations of the
data sets to vary.  The consequences of these examples will
become clearer in Section~2.7.

We incorporate correlations into the analysis by fitting the data to an
appropriate functional form $\rfit(s;a_k$) where $a_k$ are the
parameters of the function.  In the absence of undiscovered resonances,
$\rhad$ can be described by a continuous function.  A $\chi^2$ fit has
the virtue that measurements can be weighted by their experimental
errors and correlations are straightforward to include.  The previous
version of this analysis used a non-diagonal definition of $\chi^2$
constructed from the covariance matrix
$E_{ij}=\vev{\Delta\rhad^i\Delta\rhad^j}$ of the measured points
$\rhad^i$.  Unfortunately, it has been shown that if the off-diagonal
elements $E_{ij}$ scale with the measured values of $\rhad^i$ or
$\rhad^j$, the resulting fit will be biased to smaller values of
$\rhad$%
\Ref\dagostini{See G.~D'Agostini, {\it Nucl. Inst. Meth.} {\bf A346},
306 (1994) for a good discussion of this effect.}.
The bias that resulted to our previous analysis\refmark{\mls} from the
application of the incorrect technique was approximately 39\% of the
uncertainty on the final result.  We avoid the bias by defining
$\chi^2$ as follows,
$$\chi^2=\sum_i
{\left[\rhad^i-(1+\lambda_j\alpha_i)\rfit(s_i;a_k)\right]^2\over\sigma_i
^2({\rm ptp})}+\sum_j \lambda_j^2, \eqn\chidef    $$
where $\rhad^i$ is the value of $\rhad$ measured at energy $s_i$,
$\alpha_i=\sigma_i({\rm norm})/\rhad^i$ is the fractional normalization
uncertainty associated with the $i^{th}$ measurement, and $\lambda_j$
are fit parameters which are constrained to have zero mean and unit
width.  This form preserves shape information and propagates the
normalization uncertainties into the parameters of the function
$\rfit$.  For each fit, two choices of the parameters $\lambda_j$ are
investigated.  In the first case, a separate normalization parameter
$\lambda_j$ is assigned to each data set.  This choice incorporates
Type~I correlations only and makes no assumptions about correlations
between experiments.  In the second case, the normalizations of
experiments of similar age and energy region are assumed to be 100\%
correlated.   A separate normalization parameter is assigned to each
correlated group instead of each set of measurements.  This choice
includes the effects of Type~I and Type~II correlations, produces
larger error estimates (a consequence of including the Type~II
correlations), and is the one quoted as the {\it official} result.  The
difference in $\dalhad$ resulting from the two weighting schemes is
included in the parameterization uncertainty discussed below.

Equation~\dispint\ is evaluated by performing a Simpson's Rule
integration using the function $\rfit$ and the best estimate of the
parameters.  The parameter uncertainties $\delta a_k$ reflect the
point-to-point and normalization uncertainties to some extent.
Unfortunately, the process of fitting a large number of measurements
with a function of a smaller number of parameters necessarily involves
some loss of information.  The resulting uncertainty on the fitting
function at some point $W$ is usually smaller than the uncertainties on
nearby data points.  If we add {\it a priori} information to the
problem by choosing a physically motivated fitting function, the
information contained in the parameter error matrix may be appropriate.
To understand this problem better, we evaluate the uncertainty on
$\dalhadmz$ by two techniques.  In the first, the parameter
uncertainties are propagated to the calculated value of $\dalhadmz$
using the following expression which is valid for any function of the
parameters,
$$\delta^2(\dalhad)_{exp}=\sum_{k,l}{\partial(\dalhad)\over\partial
a_k} E_{kl}
{\partial(\dalhad)\over\partial a_l}, \eqn\errdef $$
where the derivatives are calculated numerically and
$E_{kl}=\vev{\delta a_k\delta a_l}$ is the parameter error matrix that
is extracted from the fitting procedure.  The second error estimate is
performed by constructing a large ensemble of data sets by shifting the
measured data points $\rhad^i({\rm meas})$ as follows,
$$\rhad^i({\rm set}~j)=\rhad^i({\rm meas})+f_{ij}^{\rm
ptp}\sigma_i({\rm ptp}) + f_{ij}^{\rm norm}\sigma_i({\rm norm}),
\eqn\shiftr $$
where the factors $f_{ij}$ are Gaussian-distributed random numbers of
unit variance.  The entire fitting and integration procedure is then
applied to each member of the ensemble.  The uncertainty on $\dalhadmz$
is determined from the central 68.3\% of the ensemble distribution.

The use of a fitting function has the problem that one may introduce
bias through the choice of parameterization.  We attempt to evaluate
this effect by varying the parameterizations as much as ingenuity and
computer time allow.  The quoted contributions to $\dalhadmz$ are those
corresponding to the best fits.  Each contribution is assigned a
parameterization uncertainty $\dalparam$ based upon the spread of
results corresponding to reasonable fits.  The parameterization
uncertainty also includes a contribution from the difference observed
in the two $\chi^2$ weighting schemes.

\section{The Data}

The approach to the evaluation of equation~\dispint\ is driven by the
form of the data themselves.  The total hadronic cross section can be
decomposed into four pieces: the hadronic continuum above
$W\equiv\roots=1$~GeV, the charged two-body final states $\pi^+\pi^-$
and $K^+K^-$ from their respective thresholds to 2.6~GeV, and hadronic
resonances (excluding charged two-body final states).  Since
equation~\dispint\ is linear in the hadronic cross section, we
decompose $\dalhad$ as follows,
$$\dalhad(q^2)=\dalhad^{\rm cont}(q^2)+\dalhad^{\pi^+\pi^-}(q^2) +
\dalhad^{\rm K^+K^-}(q^2) + \dalhad^{\rm res}(q^2), \eqn\decomp $$
where the four terms on the right-hand side correspond to the four
pieces of the hadronic cross section.

\REF\vepptwopi{CMD and OLYA Collaborations: L.M.~Barkov, \etal, {\it
Nucl. Phys.} {\bf B256}, 365 (1985).}
\REF\nas{NA7 Collaboration: S.~Amendolia, \etal, {\it Phys. Lett.} {\bf
B138}, 454 (1984).}
\REF\tof{TOF Collaboration: I.B.~Vasserman, \etal, {\it Sov. J. Nucl.
Phys.} {\bf 33}, 709 (1981).}
\REF\mtn{M2N Collaboration: G.~Cosme, \etal, {\it Phys. Rev. Lett.}
{\bf 48}, 906 (1982).}
\REF\dmopi{DM1 Collaboration: A.~Quenzer, \etal, {\it Phys. Lett.} {\bf
76B}, 512 (1978).}
\REF\mupi{$\mu\pi$ Collaboration: G.~Barbiellini, \etal, {\it Nuovo
Cim. Lett.} {\bf 6}, 557 (1973).}
\REF\dmtpi{DM2 Collaboration: D.~Bisello, \etal, {\it Phys. Lett.} {\bf
220B}, 321 (1989).}
\REF\mea{MEA Collaboration: B.~Esposito, \etal, {\it Phys. Lett.} {\bf
67B}, 239 (1977); B.Esposito, \etal, {\it Nuovo Cim. Lett.} {\bf 28},
337 (1980).}
\REF\olya{OLYA Collaboration: P.M..~Ivanov, \etal, {\it Phys. Lett.}
{\bf 107B}, 297 (1981).}
\REF\cmd{CMD Collaboration: G.V.~Anikin, \etal, IYF-83-85, August
1983.}
\REF\dmok{DM1 Collaboration: B.~Delcourt, \etal, {\it Phys. Lett.} {\bf
99B}, 257 (1981); G.~Grosdidier, \etal, LAL 80-35.}
\REF\dmtk{DM2 Collaboration: D.~Bisello, \etal, {\it Z. Phys.} {\bf
C39}, 13 (1988).}
The rationale for this decomposition is as follows. The region below
$W=1$~GeV is dominated by the $\rho$, $\omega$, and $\phi$ resonances.
The electromagnetic form factors for the processes
$\ee\to\pi^+\pi^-$\refmark{\vepptwopi-\mea} and $\ee\to
K^+K^-$\refmark{\mea-\dmtk} are measured well from threshold to
$W\simeq2$~GeV.  Resonances do not account for all of the $\pi^+\pi^-$
and $K^+K^-$ cross section in this region.  On the other hand,
essentially all other two-body and three-body final states are
associated with the resonances.  Measurements of three-pion final
states near $W=1$~GeV%
\Ref\parrour{G.~Parrour, \etal, {\it Phys. Lett.} {\bf 63B}, 357
(1976).}
show the non-resonant portion to be consistent with zero.  Similarly,
measurements of various two-body final states such as $K^0_LK^0_S$ show
small non-resonant cross sections\refmark{\cmd}.
\REF\fourpi{G. Cosme, \etal, {\it Phys. Lett.} {\bf 63B}, 349 (1976);
L.M.~Kurdadze, \etal, {\it JETP Lett.} {\bf 43}, 643 (1986);
L.M.~Kurdadze, \etal, {\it JETP Lett.} {\bf 47}, 512 (1988).}
The cross sections for four-pion final states become significant above
1~GeV but are small below that energy\refmark{\fourpi}.
The $\gamma\gamma2$ experiment%
\Ref\bacci{$\gamma\gamma2$ Collaboration: C.~Bacci, \etal, {\it Phys.
Lett.} {\bf 86B}, 234 (1979).}
\REF\lowencont{V.A.~Sidorov, {\it Proceedings of the XVIII$^{\rm th}$
Conference on High Energy Physics}, ed. N.N.~Bogolubov, Tbilisi, 1976.}
at the ADONE storage ring at Frascati has measured the hadronic cross
section ratio for three or more hadron final states, $\rhadt$ from
$W=1.42$~GeV to $W=3.09$~GeV.  They have also presented several points
from 1~GeV to 1.4~GeV
that are composed of various multipion cross sections from Novosibirsk
and Orsay\refmark{\fourpi,\parrour,\lowencont} and are claimed to
approximate $\rhadt$.
Measurements beginning at $W=2.6$~GeV by the MARK~I%
\Ref\mki{MARK~I collaboration: J.L.~Siegrist, \etal, {\it Phys. Rev.}
{\bf D26}, 969 (1982); J.L.~Siegrist, SLAC-Report No.~225, October
1979.},
DASP%
\Ref\dasp{DASP Collaboration: R.~Brandelik, \etal, {\it Phys. Lett.}
{\bf 76B}, 361 (1978); R.~Brandelik, \etal, {\it Z. Phys.} {\bf C1},
233 (1979); H.~Albrecht, \etal, {\it Phys. Lett.} {\bf 116B}, 383
(1982).},
PLUTO%
\Ref\pluto{PLUTO Collaboration: J.~Burmeister, \etal, {\it Phys. Lett.}
{\bf 66B}, 395 (1977); Ch.~Berger, \etal, {\it Phys. Lett.} {\bf 81B},
410 (1979).},
and Crystal Ball%
\Ref\cbcthr{Crystal Ball Collaboration: A.~Osterheld, \etal,
SLAC-PUB-4160, December 1986.}
Collaborations claim to measure the entire cross section.  We therefore
conclude that $\rhad$ is well approximated below $W_1=2.6$~GeV by a sum
of the $\pi^+\pi^-$ and $K^+K^-$ contributions from threshold to $W_1$
(where they are much smaller than $\rhadt$); the $\rhadt$ measurements
from 1~GeV to $W_1$; and the $\rho$, $\omega$ and $\phi$ resonances
where the hadronic widths are adjusted to remove the $\pi^+\pi^-$ and
$K^+K^-$ final states that are already included explicitly.  Note that
the several broad $\ee$ resonances between the $\phi(1020)$ and
$W=2$~GeV are implicitly contained in the two-body or $\rhadt$
categories.  Since the $\pi^+\pi^-$ and $K^+K^-$ cross sections are
very small at $W_1$, the $\rhadt$ and total continuum $\rhad$
measurements should be continuous at this point.

\REF\cb{Crystal Ball Collaboration: Z.~Jakubowski, \etal, {\it Z.
Phys.} {\bf C40}, 49 (1988); C.~Edwards, \etal, SLAC-PUB-5160, January
1990.}
\REF\lena{LENA Collaboration: B.~Niczyporuk, \etal, {\it Z. Phys.} {\bf
C15}, 299 (1982).}
\REF\cleo{CLEO Collaboration: R.~Giles, \etal, {\it Phys. Rev.} {\bf
D29}, 1285 (1984).}
\REF\cusb{CUSB Collaboration: E.~Rice, \etal, {\it Phys. Rev. Lett.}
{\bf 48}, 906 (1982).}
\REF\desyheid{DESY-Heidelberg Collaboration: P.~Bock, \etal, {\it Z.
Phys.} {\bf C6}, 125 (1980).}
\REF\chetkuhn{K.G.~Chetyrkin and J.H.~Kuhn, TTP-94-12, September 1994.}
\Picture\CBMKI
\file={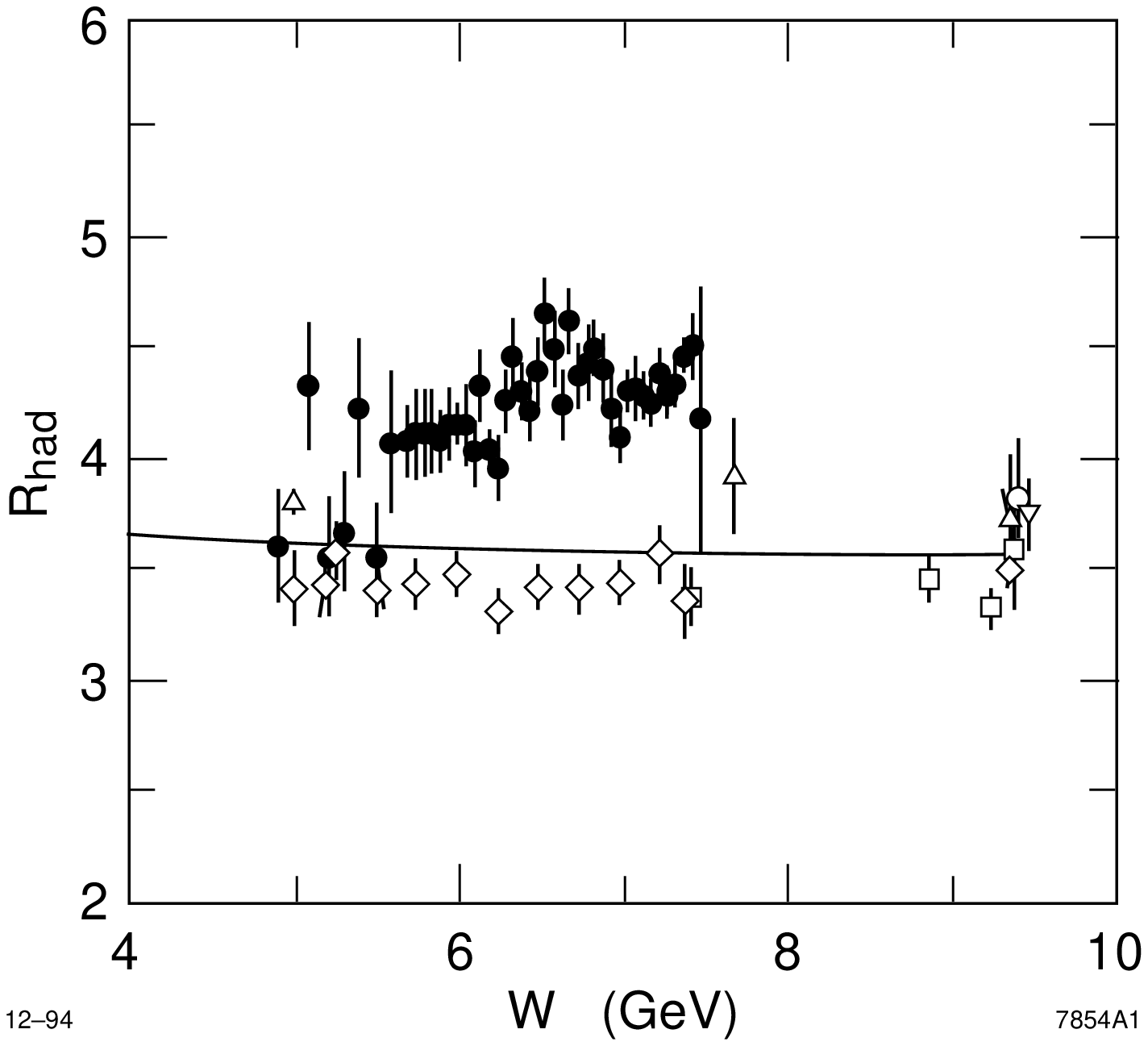}
\width=4.0in
\caption={
The $\rhad$ measurements of the MARK~I\refmark{\mki} (solid squares),
PLUTO\refmark{\pluto} (open triangles), Crystal
Ball\refmark{\cb} (open diamonds), LENA\refmark{\lena} (open squares),
DASP\refmark{\dasp} (open inverted triangle), and
DESY-Heidelberg\refmark{\desyheid} (open circle) Collaborations in the
region between $W=5$~GeV and $W=9.4$~GeV.  The error bars include
point-to-point uncertainties only.  A recent QCD
calculation\refmark{\chetkuhn} which includes quark mass effects is
shown as a solid line for $\alpha_s(M_Z^2)=0.125$.}
\savepicture\piCBMKI

At center-of-mass energies larger than $W_1$, many measurements of the
hadronic continuum and resonances exist.  The only precise measurement
in the region between 2.6~GeV and 5.0~GeV is a single data point just
below charm threshold at $W=3.670$~GeV by the Crystal Ball
Collaboration\refmark{\cbcthr}.  This measurement has a normalization
uncertainty of 7\%.  Since the next most precise measurements in the
region below 5~GeV have normalization uncertainties of 15\%, this
measurement represents an important constraint on the magnitude of the
cross section in the entire region.  The region above charm threshold
from $W=3.77$~GeV to $W=5.0$~GeV is complicated and not well measured.
The MARK~I, DASP, PLUTO and Crystal Ball Collaborations all observe an
enhancement beyond the expected threshold shape.  The DASP data show
three resolved resonances.  The MARK~I and PLUTO data are consistent
with the DASP data but do not cleanly resolve the resonances.  The
Crystal Ball measurements are somewhat smaller than the older ones and
do not resolve the second resonance (which appears as a broad
shoulder).  We choose to follow the Particle Data Group and recognize
the DASP resonances: $\psi(4040)$, $\psi(4160)$, and $\psi(4415)$.  The
$\psi$ family therefore consists of six states.

Between 5~GeV and 10.4~GeV, the MARK~I, DASP, PLUTO, Crystal
Ball\refmark{\cb}, LENA\refmark{\lena}, CLEO\refmark{\cleo},
CUSB\refmark{\cusb}, and DESY-Heidelberg\refmark{\desyheid}
Collaborations have published $\rhad$ measurements which are plotted in
Figure~\CBMKI.  The error bars include only point-to-point
uncertainties.  The recently published Crystal Ball measurements have a
systematic normalization uncertainty of 5.2\%.  The other measurements
have normalization uncertainties in the range 6.8-10\%.  The data are
also compared with the recent QCD prediction of Chetyrkin and
Kuhn\refmark{\chetkuhn} which includes quark mass effects.  At
$W=5$~GeV, the MARK~I data are consistent with other measurements.  As
$W$ increases,  they show a systematic increase in $\rhad$ and suggest
the presence of a structure near 6.6~GeV.  Including the quoted 10\%
normalization uncertainty, the MARK~I data are larger than the more
precise measurements by approximately
two standard deviations.  The reader is reminded that first generation
detectors like MARK~I, DASP, and PLUTO were small acceptance devices
that necessarily involved large acceptance corrections without the
benefit of good event structure modelling.  After acceptance
corrections and a $\tau$-lepton subtraction, the MARK~I group observed
that two-charged-prong events constituted nearly 20\% of the hadronic
cross section of $R$ at $W=7$~GeV.  This is about 1.5 times the
two-prong rate due to $\tau^+\tau^-$ final states and three times the
rate that is predicted%
\ref{P.~Burrows and Y.~Ohnishi, private communication.}
by the JETSET~7.3 Monte Carlo program%
\ref{T.~Sj\"ostrand, CERN-TH-6488-92, May 1992.}.
While this may not be wrong, we choose to exclude data from the first
generation experiments when more modern results are available.  Such
data are available above charm threshold.  Unfortunately, we are
constrained to use very old continuum measurements below charm
threshold.

\centerline{\piCBMKI}

The Particle Data Group lists six $\Upsilon$ family resonances between
9.4~GeV and 11~GeV.  All are included in the resonance contribution.

\REF\hrs{HRS Collaboration: D.~Bender, \etal, {\it Phys. Rev.} {\bf
D31}, 1 (1985).}
\REF\mac{MAC Collaboration: E.~Fernandez, \etal, {\it Phys. Rev.} {\bf
D31}, 1537 (1985).}
\REF\cello{CELLO Collaboration: H.-J.~Behrend, \etal, {\it Phys. Lett.}
{\bf B183}, 407 (1987).}
\REF\jade{JADE Collaboration: W.~Bartel, \etal, {\it Phys. Lett.} {\bf
129B}, 145 (1983); {\it Phys. Lett.} {\bf B160}, 337 (1985);
B.~Naroska, \etal, {\it Phys. Rept.} {\bf 148}, 67 (1987).}
\REF\markj{MARK~J Collaboration: B.~Adeva, \etal, {\it Phys. Rev.} {\bf
D34}, 681 (1986).}
\REF\tasso{TASSO Collaboration: R.~Brandelik, \etal, {\it Phys. Lett.}
{\bf 113B}, 499 (1982); M.~Althoff, \etal, {\it Phys. Lett.} {\bf
B138}, 441 (1984).}
Above b-quark threshold, a number of $\rhad$ measurements have been
carried out by the PEP and PETRA experiments\refmark{\hrs-\tasso}.
However at energies above $W=34$~GeV, $Z$-$\gamma$ interference becomes
significant.  We therefore use only those measurements in the region
$W\le34$~GeV where the correction for electroweak interference is less
than 1\%.

We expect that $\rhad$ is well described by perturbative QCD in the
region above b-quark threshold.  This implies that the world average
value of the strong coupling constant $\alpha_s(M_Z^2)$ compiled by the
Particle Data Group
\refmark{\pdg} provides a precise measurement of $\rhad$ at $W=M_Z$.
Since possible anomalies in the $Z$ lineshape would bias the
determination of $\alpha_s(M_Z^2)$ from the lineshape parameters, we
exclude the $Z$ lineshape information from the Particle Data Group
average.  Additionally, since we explicitly include the PEP/PETRA
$\rhad$ measurements in our fit (which uses perturbative QCD to
describe the PEP/PETRA energy region), they are also excluded from the
PDG average yielding the following value,
$$    \alpha_s(M_Z^2)=0.116\pm0.005. \eqn\strongcc     $$
To convert $\alpha_s(M_Z^2)$ into a determination of $\rhad(M_Z)$, we
use the third-order QCD expression%
\Ref\rthird{S.G.~Gorishny, A.L.~Kataev, and S.A.~Larin, {\it Phys.
Lett.} {\bf B259}, 144 (1991); L.R.~Surgladze, M.A.~Samuel, {\it Phys.
Rev. Lett.} {\bf 66}, 560 (1991); Erratum: ibid, 2416.},
$$\eqalign{\rqcd(s)=3\sum_f&\ Q_f^2\beta_f{(3-\beta_f^2)\over2}\cr
&\cdot\left\lbrace 1+\left[{\alpha_s(s)\over\pi}\right]
+r_1\left[{\alpha_s(s)\over\pi}\right]^2
+r_2\left[{\alpha_s(s)\over\pi}\right]^3\right\rbrace,\cr} \eqn\rqcdeq
$$
where: $Q_f$ is the final state fermion charge,
$\beta_f=\sqrt{1-4m_f^2/s}$ is the fermion velocity in the $\ee$
center-of-mass frame ($m_f$ is the fermion mass), and the coefficients
are functions of the number of active flavors $N_f$,
$$\eqalign{r_1=&\ 1.9857-0.1153N_f\cr
r_2=&\ -6.6368-1.2002N_f-0.0052N_f^2-1.2395{\left(\sum
Q_f\right)^2\over 3\sum Q_f^2}.\cr} \eqn \coeffdef        $$
The resulting value of $\rhad(M_Z)$ is,
$$\rhad(M_Z)=3.807\pm0.006.    \eqn\rhadmz  $$

The following three sections of this chapter describe the evaluation
of: the continuum contribution $\dalhad^{\rm cont}$, the contributions
of the charged two-body final states $\dalhad^{\pi^+\pi^-}$ and
$\dalhad^{K^+K^-}$, and the resonance contribution $\dalhad^{\rm res}$.

\section{The Hadronic Continuum}

The first step in the evaluation of equation \dispint\ for the hadronic
continuum is to formulate a suitable (piecewise-continuous)
parameterization
$\rfit(s;a_k)$.  We choose to use the perturbative QCD expression given
in equation~\rqcdeq\ in the region $W\ge15$~GeV and an empirical
parameterization in the region 1~GeV$\leq W<15$~GeV.  In the high
energy region, the only free parameter is $\alpha_s(M_Z^2)$ which is
evolved to other scales numerically using the Runge-Kutta method
applied to the order-$\alpha_s^4$ renormalization group equation%
\ref{See Ref.~\pdg\ and errata.}.

In the portions of the low energy region that are measured well,
polynomials are used to parameterize $\rhad(W)$.  To ensure that the
function is continuous across several points $W_a$, the polynomials are
constructed in $x_a\equiv W-W_a$ and the zeroth order terms are
excluded,
$$P_n^a(x)\equiv\sum_{i=1}^nd^a_ix_a^i,  \eqn\pdef $$
where $a$ is a label to distinguish different regions.  Separate
polynomials are used to describe the following regions: 1~GeV$\leq
W\leq1.9$~GeV (labelled region s), 1.9~GeV$<W\leq3.6$~GeV (labelled
region c), and 5.0~GeV$<W\leq10.4$~GeV (labelled region b).  Although a
single, large-order polynomial is adequate to describe the data between
$W=1$~GeV and charm threshold at 3.6~GeV, the data show a distinct
shape change near $W=1.9$~GeV (where the four-pion cross section is
becoming small).  It was possible to obtain better fits by introducing
an additional polynomial to describe the region from 1~GeV to 1.9~GeV.
A comparison of the two possible forms is used to assess the
parameterization sensitivity of the final result.

Since there are no measurements of the continuum $\rhad$ in the b-quark
and c-quark threshold regions (published measurements include a mixture
of continuum and resonances), it is necessary to extrapolate the form
of $\rhad$ from 3.6~GeV to 5.0~GeV and from 10.4~GeV to 15~GeV with
functions that are physically motivated.  In the case of the charm
threshold region, the DASP Collaboration has published (in graphical
form) the shape of the continuum that was preferred by their fit to the
$\psi(4040)$, $\psi(4160)$, and $\psi(4415)$ resonances.  The function
which characterizes the shape of the threshold, $f_{DASP}(W)$, does not
increase as sharply as the free-quark threshold factor
$\beta(3-\beta^2)/2$ but increases more rapidly than the $\beta^3$
threshold factor for pointlike scalar particles.  To construct the
function $\rfit$, all three possibilities are used for the c-quark
threshold and the two extreme possibilities are used for the b-quark
threshold,
$$f_c(W)=\cases{\beta(3-\beta^2)/2\cr f_{DASP}(W)\cr \beta^3\cr} \qquad
f_b(W)=\cases{\beta(3-\beta^2)/2\cr \beta^3,\cr} \eqn\fdef    $$
where the c- and b-quark masses are taken to be the $D$ and $B$ meson
masses, respectively.  The actual size of the charm-associated step in
$\rhad$, $\Delta R_c$ is left as a free parameter.  The size of the
bottom-associated step in $\rhad$ is constrained to be the difference
between the value of the fit function at $W=10.4$~GeV and the value of
the QCD portion at $W=15$~GeV, $\Delta R_b\equiv
R_{QCD}(15)-R_{fit}(10.4)$.

\Picture\CONT
\file={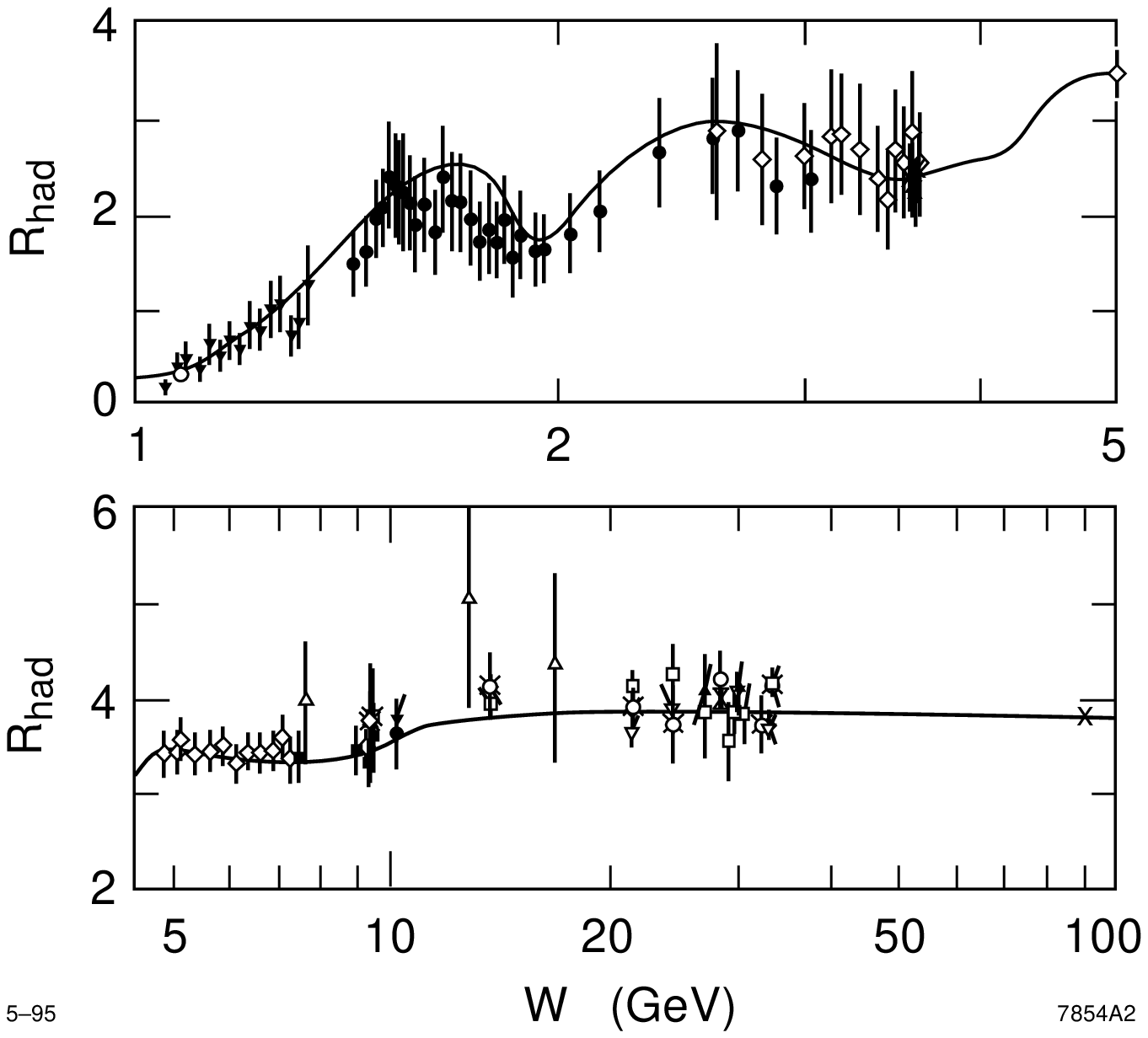}
\height=3in
\width=5in
\caption={\noindent
The continuum $\rhad$ measurements including normalization uncertainties.
The entries in the region below charm threshold consist of a compilation
of low energy exclusive cross sections\refmark{\fourpi,\parrour,\lowencont}
(solid inverted trangles) and the measurements of the
$\gamma\gamma2$\refmark{\bacci} (solid dots), Mark~I\refmark{\mki}
(open diamonds), DASP\refmark{\dasp} (X's), Crystal Ball\refmark{\cbcthr}
(solid square), and PLUTO\refmark{\pluto} (solid diamond) Collaborations.
The entries in the region between charm and bottom thresholds are the
measurements of Crystal Ball\refmark{\cb} (open diamonds),
PLUTO\refmark{\pluto} (open triangles), LENA\refmark{\lena} (solid
squares), DASP\refmark{\dasp} (diamond-X overlay),
DESY-Heidelberg\refmark{\desyheid} (square-X overlay), CUSB\refmark{\cusb}
(solid dot), and CLEO\refmark{\cleo} (solid inverted triangle)
Collaborations.  The entries in the region between above bottom threshold
and below the $Z$ pole
are the measurements of CELLO\refmark{\cello} (open diamonds),
PLUTO\refmark{\pluto} (open triangles), JADE\refmark{\lena}
(open squares), Mark-J\refmark{\markj} (open inverted triangles),
TASSO\refmark{\tasso} (circle-X overlay), HRS\refmark{\hrs} (open circle),
and MAC\refmark{\mac} (X) Collaborations.  The fit used to to evaluate the
central value of $\dalhad^{\rm cont}$ is shown as the solid curve.}
\savepicture\piCONT

The actual form of the fitting function is given by the following
expression,
$$R_{fit}(W)=\cases{R_0+P_{N_s}^s(W-1.0),&$1\leq W\leq 1.9$\cr
  R_{fit}(1.9)+P_{N_c}^c(W-1.9),&$1.9< W\leq3.6$\cr
  R_{fit}(3.6)+\Delta R_cf_c(W),&$3.6<W\leq5.0$\cr
  R_{fit}(5.0)+P_{N_b}^b(W-5.0),&$5.0<W\leq10.4$\cr
  R_{fit}(10.4)+\Delta R_bf_b(W),&$10.4<W<15.0$\cr
  R_{QCD}(W),&$15\leq W$\cr} \eqn\rcont $$
where $R_0$, the value of $\rhad$ at $W=1$~GeV, is a free parameter and
the order of the polynomials is varied from 1 to 7.  The $\chi^2$ is
constructed from equation~\chidef\ assuming that normalization
uncertainties are completely correlated in four groups: the 20\%
uncertainties of the lowest energy
measurements\refmark{\bacci-\lowencont} (1.0~GeV$<W<3.09$~GeV), the
15-20\% uncertainties of the MARK~I, DASP, and PLUTO
measurements\refmark{\mki-\pluto} (2.6~GeV$<W<4.9$~GeV), the 5-10\%
uncertainties of the measurements\refmark{\cb-\desyheid} between charm
and bottom thresholds (the Crystal Ball measurement at 3.670~GeV is
treated as a member of the higher-energy Crystal Ball set), and the
1.7-7.0\% uncertainties of the PEP and PETRA
experiments\refmark{\hrs-\tasso} above bottom threshold.  Each fit is
repeated with a separate normalization parameter for the 20 sets of
data in the analysis.

The data are corrected for electroweak interference and incomplete
vacuum polarization corrections (see Appendix~A) before the fitting
procedure is applied.  In the course of varying the orders of the
polynomials and the number of normalization parameters, the number of
free parameters varies from 12 to 44.  The fit quality does not improve
substantially when the number of parameters exceeds 14.   The data and
the result of the fit used to to evaluate the central value of
$\dalhad^{\rm cont}$ are shown in Fig.~\CONT.  The error bars include
the point-to-point and the normalization uncertainties.  The fit
quality is reasonable ($\chi^2/{\rm dof}=110/100$).
\endpage
\centerline{\piCONT}
\endpage

The various hypotheses for $\rfit$ are used to evaluate the integral in
equation~\dispint\ from $s_0=1$~GeV$^2$ to $\infty=10^6$~GeV$^2$.
Although the singularity in the integrand is formally well controlled,
digital computers are famous for their inability to understand
formalities.  We have therefore recast equation~\dispint\ into a form
which is more suitable for electronic evaluation,
$$\eqalign{\dalhad(q^2)={\alpha_0q^2\over3\pi}&\Biggl\lbrace
{\rfit(q^2)\over q^2}{\rm ln} \left[{q^2-s_0\over
s_0}\right]-\int_{s_0}^{q^2-\Delta} ds {\rfit(s)-\rfit(q^2)\over
s(s-q^2)}\cr &-{\partial\rfit\over\partial s} \Bigr\vert_{q^2}{\rm
ln}\biggl[{q^2+\Delta\over q^2-\Delta}\biggr]
 -\int_{q^2+\Delta}^\infty ds{\rfit(s)-\rfit(q^2)\over
s(s-q^2)}\Biggr\rbrace,\cr} \eqn\dispfit $$
where we have assumed that $\rfit$ is well approximated by a linear
expansion over the interval $q^2-\Delta<s<q^2+\Delta$ (in practice, we
use $\Delta=0.5$~GeV$^2$).  The evaluation of equation~\dispfit\
requires that $\alpha_s$ be evolved to scales larger than the t-quark
mass.  For this purpose, the top quark mass is assumed to be 172.3~GeV
which is the $\overline{{\rm MS}}$ mass corresponding to a pole mass of
180~GeV.

The contribution of the hadronic continuum to $\dalhadmz$ is found to
be fairly insensitive to the form of $\rfit$ and the number of
normalization parameters used.  The central value of $\dalhadmz$
corresponds to the best estimate of the parameters of the function
which uses: the DASP shape for the c-quark-threshold, the free-quark
shape for the b-quark-threshold, the values (2,3,3) for
($N_b$,$N_c$,$N_s$) and four normalization parameters.  The maximum
deviation from this value occurs when $N_b=1$ and four (instead of 20)
normalization parameters are used (the deviation is insensitive to the
choice of threshold functions).  The size of the maximum deviation is
taken as an estimate of the parameterization uncertainty.  The
experimental uncertainty given by equation \errdef\ is found to be in
excellent agreement with the estimate derived from an ensemble of 500
fluctuated data sets.  The resulting contribution to $\dalhadmz$ is
$$\dalhad^{\rm cont}(M_Z^2)=0.022106\pm0.000366({\rm
exp})\pm0.000196({\rm param}). \eqn\contans $$

\Picture\INTUNC
\file={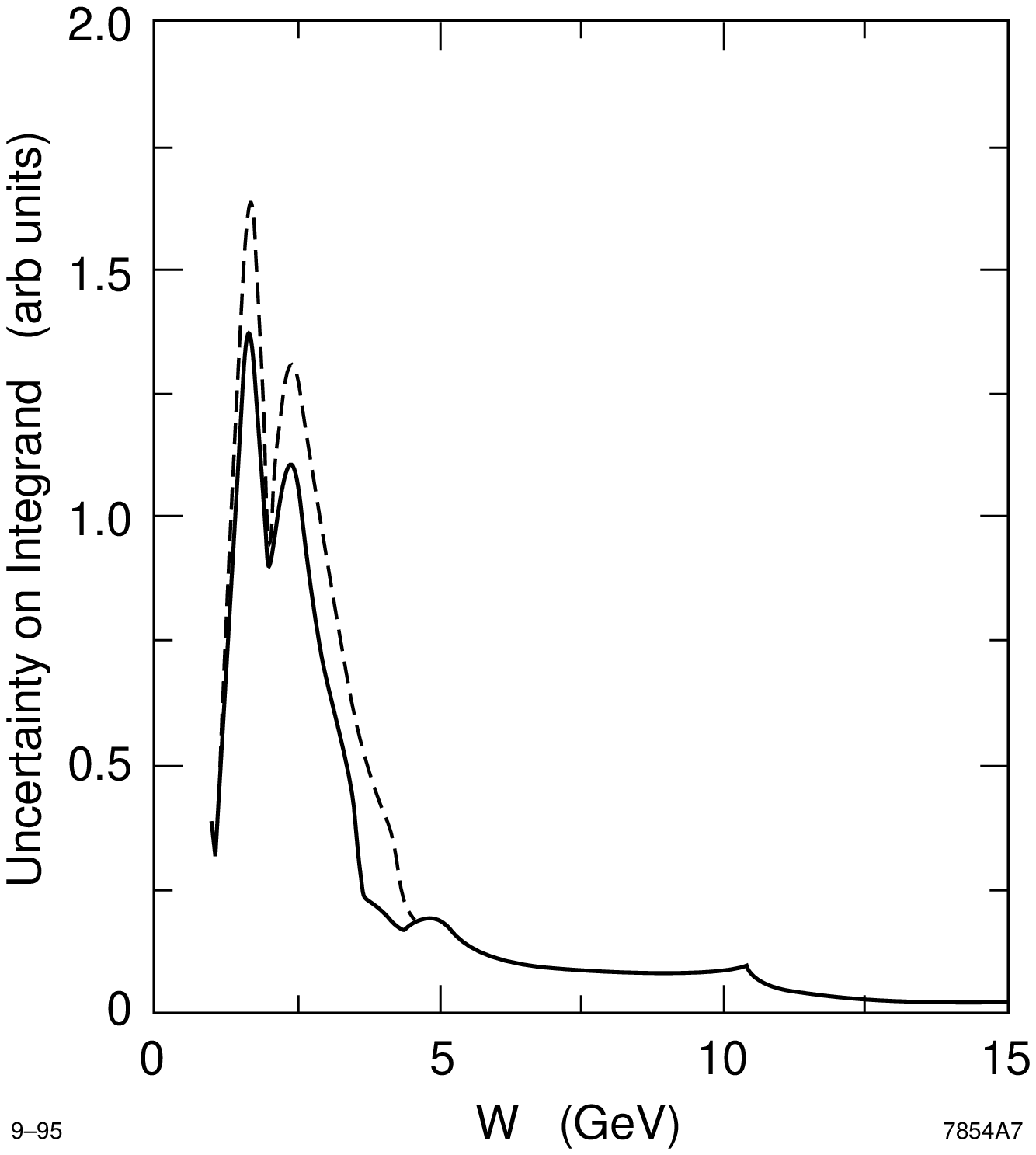}
\height=3in
\width=3.5in
\caption={\noindent
The uncertainty on the integrand of the dispersion integral (integrated
over $W$ rather than $s$) in arbitrary units.  The dashed curve shows
the uncertainty before the Crystal Ball data point is included in the
fit and the solid curve shows the uncertainty after its inclusion.}
\savepicture\piINTUNC

\centerline{\piINTUNC}

This result differs from our previous result\refmark{\mls} by
+0.000678.  Most of the difference is caused by inclusion of Crystal
Ball data point at 3.670~GeV (+0.000575).  The remaining difference is
due to the use of the five-flavor definition of $\dalhad$ (+0.000059),
a change in the value of $\alpha_s(M_Z^2)$ used as input ($-$0.000051),
and the change to the unbiased fitting technique (+0.000095).  The
inclusion of the Crystal Ball point pulls the fit to somewhat larger
values of $\rhad$ and substantially constrains the normalization in the
charm threshold region.  The Mark~II and $\gamma\gamma2$ data span a
large energy region and constrain the shape of $\rfit(W)$ down to
$W=1.4$~GeV.  The effect of the single precise point is therefore
propagated to to smaller energies.  This type of effect is illustrated
in Figure~\shape(a) and is demonstrated in Figure~\INTUNC\
which displays the uncertainty on the integrand of the
$W$-space dispersion integral in arbitrary units
\ref{The $W$-space integrand $I^\prime(W)$ is related to the $s$-space
integrand $I(s)$ by the following simple expression
$I^\prime(W)=2WI(W^2)$}.  The uncertainty is calculated using
equation~\errdef\ (with $\dalhad$ replaced by $\rfit$) to estimate the
uncertainty on $\rfit(W)$ at each energy point.  The dashed curve shows
the uncertainty before the Crystal Ball data point
is included in the fit and the solid curve shows the uncertainty after
its inclusion.  Note that the overall uncertainty on $\dalhad^{\rm
cont}$ is dominated by the poor precision of the data in the 1~GeV to
3.5~GeV region.

\section{The $\pi^+\pi^-$ and $K^+K^-$ Final States}

The processes $\ee\to\pi^+\pi^-$ and $\ee\to K^+K^-$ are described by
the electromagnetic form factors, $F_\pi(s)$ and $F_K(s)$, which are
related to the hadronic cross section ratio $\rhad$ for each process as
follows,
$$\rhad^{\pi^+\pi^-}(s) = {1\over4}|F_\pi(s)|^2\beta_\pi^3,\qquad
\rhad^{K^+K^-}(s) = {1\over4}|F_K(s)|^2\beta_K^3, \eqn\formfac $$
where $\beta_\pi$ and $\beta_K$ are the velocities of the final state
particles in the $\ee$ center-of-mass frame.  It is clear that
measurements of the form factors are equivalent to measurements of
$\rhad$.

\Picture\PIPI
\file={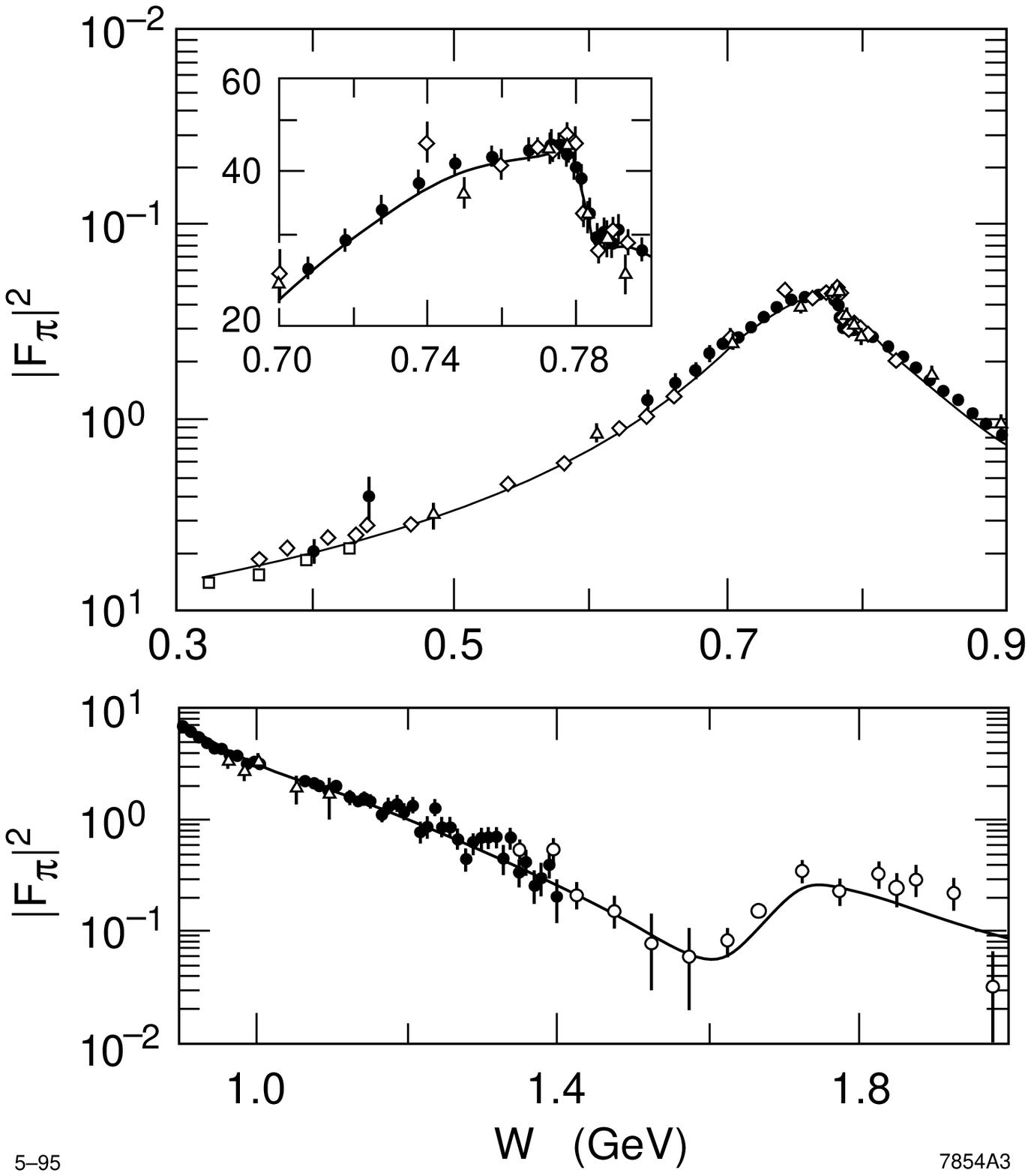}
\height=3in
\width=5.2in
\caption={\noindent
Measurements of $|F_\pi(W)|^2$ by the OLYA\refmark{\vepptwopi}
(solid dots), CMD\refmark{\vepptwopi} (open diamonds), TOF\refmark{\tof}
(solid triangles), NA7\refmark{\nas} (open squares),
$\mu\pi$\refmark{\mupi} (solid squares), MEA\refmark{\mea}
(solid diamonds), DM1\refmark{\dmopi} (open triangles), and
DM2\refmark{\dmtpi} (open circles) Collaborations are compared with
the best fit which is shown as a solid line.  The error bars include
normalization uncertainties.}
\savepicture\piPIPI

Measurements of the square of the pion form factor $|F_\pi|^2$ have
been performed by the OLYA\refmark{\vepptwopi},
CMD\refmark{\vepptwopi}, TOF\refmark{\tof}, NA7\refmark{\nas},
$\mu\pi$\refmark{\mupi}, MEA\refmark{\mea}, M2N\refmark{\mtn},
DM1\refmark{\dmopi}, and DM2\refmark{\dmtpi} Collaborations and are
shown in Fig.~\PIPI.  The error bars include the normalization
uncertainties which range from about 2\% in the region around the
(dominant) $\rho$ resonance to about 12\% at $W\simeq2$~GeV.

\REF\gs{G.J.~Gounaris and J.J.~Sakurai, {\it Phys. Rev. Lett.} {\bf
21}, 244 (1968).}
\REF\kno{T.~Kinoshita, B.~Nizic and Y.~Okamoto, {\it Phys. Rev.} {\bf
D31}, 2108 (1985).}
The data are first corrected for incomplete vacuum polarization
corrections as described in Appendix~A.  They are then fit to a
function which is a sum of the Gounaris-Sakurai form\refmark{\gs} used
by  Kinoshita, Nizic, and Okamoto\refmark{\kno} and three resonances,
$$F_\pi(s)={A_1-m_\pi^2A_2\over A_1+A_2q^2+f(s)} +
\sum_{n=1}^3{B_ne^{iC_n}m_n^2\over s-m_n^2+im_n\Gamma_n}, \eqn\pipifcn
$$
where: $A_1$ and $A_2$ are free parameters; $m_\pi$ is the pion mass;
$q$ and $f(s)$ are defined as follows,
$$\eqalign{q\equiv&\ \sqrt{s/4-m_\pi^2}\cr
f(s)\equiv&\
{1\over\pi}\left[m_\pi^2-{s\over3}\right]+{2q^3\over\pi\sqrt{s}} {\rm
ln} \left[{\sqrt{s}+2q\over2m_\pi}\right]-i{q^3\over\sqrt{s}};\cr}
\eqn\pipiaux $$
and where $m_n$, $\Gamma_n$, $B_n$, and $C_n$ are the mass, width,
amplitude, and phase of each resonance.  The mass and width of the
first resonance are set to those of the $\omega(782)$.  All other
parameters (12 in total) are allowed to vary.  The $\chi^2$ function is
constructed assuming that all normalization uncertainties are 100\%
correlated (one normalization parameter) and that the normalizations
are uncorrelated (seven normalization parameters).  As in the case of
the continuum, the two fits give nearly identical results but the error
estimate is larger when only one normalization parameter is used.  The
result of the single-normalization-parameter-fit is shown as a solid
line in Fig.~\PIPI.  The fit preferred a resonance of width 0.44~GeV at
mass 1.15 GeV and a second resonance of width 0.18~GeV at mass
1.71~GeV.  The fit quality is found to be good ($\chi^2/{\rm
dof}=138.3/127$).

To evaluate the sensitivity of the result to the parameterization, the
complete function used by the authors of Ref.~\kno\ was also fit to the
data.  This function did not fit the newest (large $W$) data from DM2
as well as our chosen form ($\chi^2/{\rm dof}=201.5/132$).  Both
functions were used to evaluate equation \dispint\ from $s=4m_\pi^2$ to
$s=4$~GeV$^2$ (where $|F_\pi|^2$ is measured to be very small).  We
find the $\pi^+\pi^-$ contribution to $\dalhadmz$ to be
$$\dalhad^{\pi^+\pi^-}(M_Z^2)=0.003240\pm0.000057({\rm
exp})\pm0.000169({\rm param}). \eqn\pipians $$
The two techniques for the estimation of the experimental uncertainty
(discussed in Section~2.1) yield consistent results.

The result given in equation~\pipians\ differs from our previous
result\refmark{\mls} by +0.000153.  The difference is due entirely to
the use of the unbiased fitting technique and represents the largest
problem found with the older technique.

\centerline{\piPIPI}

\Picture\KPKM
\file={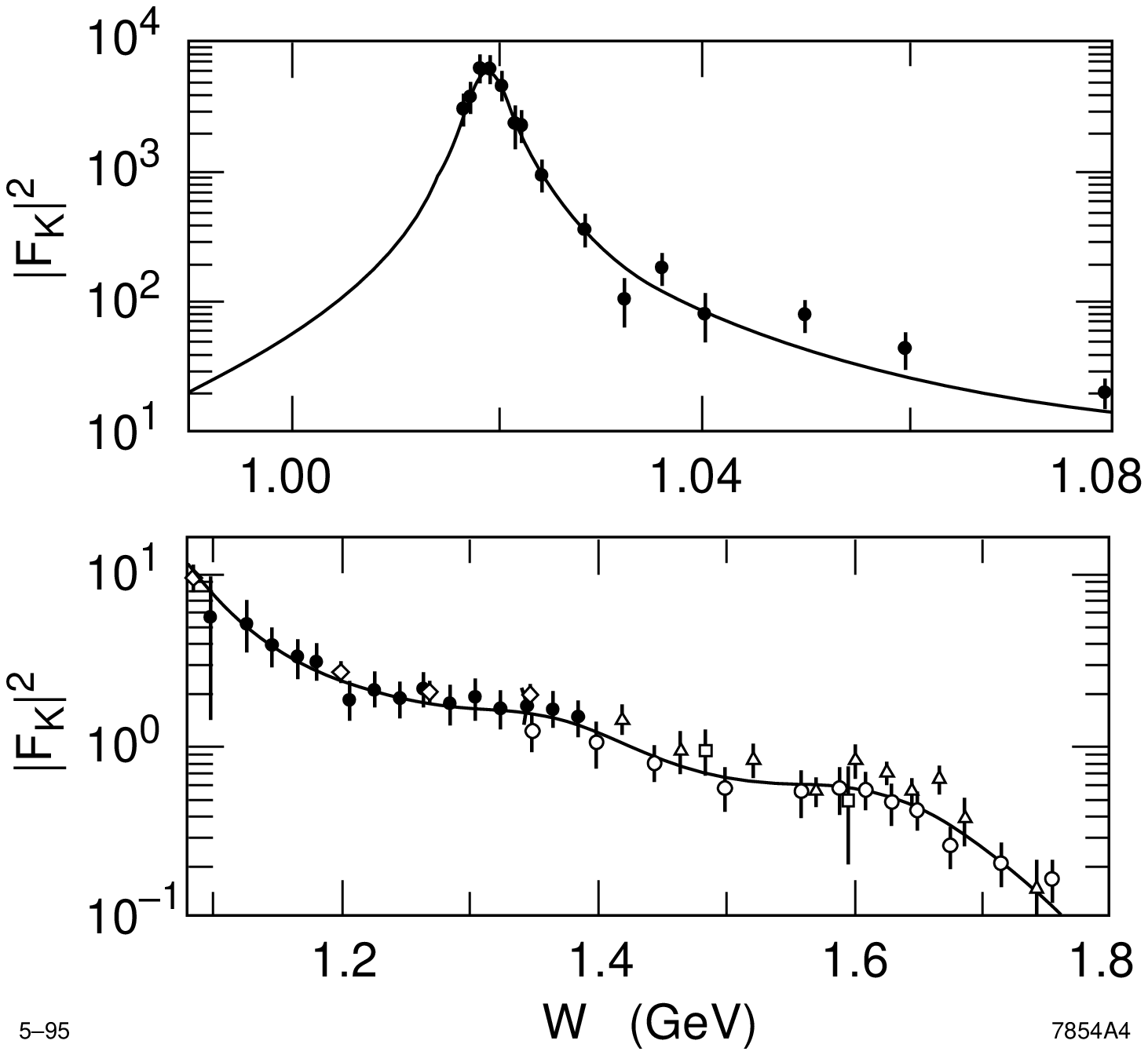}
\height=3in
\width=5.2in
\caption={\noindent
Measurements of $|F_K(W)|^2$ by the OLYA\refmark{\olya} (solid dots),
CMD\refmark{\cmd} (open diamonds), MEA\refmark{\mea} (open squares),
DM1\refmark{\dmok} (open triangles), and DM2\refmark{\dmtk}
(open circles) Collaborations are compared with the best fit which is
shown as a solid line.  The error bars include normalization
uncertainties.}
\savepicture\piKPKM

Measurements of the square of the kaon form factor $|F_K|^2$ have been
performed by the OLYA\refmark{\olya}, CMD\refmark{\cmd},
MEA\refmark{\mea}, DM1\refmark{\dmok}, and DM2\refmark{\dmtk}
Collaborations and are shown in Fig.~\KPKM.  The data span the
$\phi(1020)$ resonance and continue to $W=1.8$~GeV where
$\rhad^{K^+K^-}$ is less than 0.01.  The normalization uncertainty on
the CMD measurements is 6\%.  The other groups do not report
normalization uncertainties.  Early $|F_\pi|^2$ measurements suffered
from the same problem of unreported normalization uncertainties.  A bit
of historical research shows that the normalization uncertainties were
usually not included in the point-to-point errors.  We therefore
arbitrarily assign a 20\% systematic normalization uncertainty to all
unreported cases.  The data and total uncertainties are shown in
Fig.~\KPKM.

\centerline{\piKPKM}

The data are fit to a function which is a sum of a Breit-Wigner
resonance with an energy-dependent width for the $\phi$ and four
resonances,
$$F_K(s)={A_1\over s-m_\phi+im_\phi\Gamma_\phi(s)} +
\sum_{n=1}^4{B_ne^{iC_n}\over s-m_n^2+im_n\Gamma_n}, \eqn\kpkmfcn $$
where: $A_1$ is the amplitude of the $\phi$; $m_\phi$ is the mass of
the $\phi(1020)$; $m_n$, $\Gamma_n$, $B_n$, and $C_n$ are the mass,
width, amplitude, and phase of the resonances.  The energy-dependent
width $\Gamma_\phi(s)$ is assumed to consist of contributions from the
$K^+K^-$, $K_LK_S$, and $3\pi$ final states,
$$\Gamma_\phi(s)=\Gamma_\phi^0\left\lbrace{\sqrt{s}\over
m_\phi}\left[0.497{\beta_+^3(s)\over\beta_+^3(m_\phi^2)} +
0.347{\beta_0^3(s)\over\beta_0^3(m_\phi^2)}\right]+0.156G^\phi_{3\pi}(s)
 \right\rbrace, \eqn\gamphi $$
where: $\Gamma_\phi^0$ is the nominal value\refmark{\pdg} of the $\phi$
width, $\beta_+(s)=\sqrt{1-4m_{K^+}^2/s}$ is the velocity of the
charged kaon, $\beta_0(s)=\sqrt{1-4m_{K^0}^2/s}$ is the velocity of the
neutral kaon, and $G^\phi_{3\pi}(s)$ is a function which is normalized
to unity at $s=m_\phi^2$ and is proportional to the decay rate for
$\phi\to3\pi$ assuming $\rho\pi$ dominance%
\Ref\vvs{The form of the interaction Lagrangian for two vector fields
(field tensors $F_1^{\mu\nu}$ and $F_2^{\mu\nu}$) and a pseudoscalar
field $\phi$ is assumed to be ${\cal L}\sim F_1^{\mu\nu}\widetilde
F_{2\mu\nu}\phi/m_1$ where $m_1$ is the mass of the parent particle.}.

The masses and widths of the first two resonances were set to those of
the $\rho(770)$ and $\omega(782)$.  Following the procedure of
Ref.~\dmtk, the amplitude ratios $B_1/A_1$ and $B_2/A_1$ were
constrained to the measured values and the phases were set to zero.
The mass, width, and amplitude of the $\phi$ were allowed to vary.  The
masses, widths, amplitudes, and phases of two larger mass resonances
were free parameters.  The $\chi^2$ function was constructed with the
assumptions that all normalization uncertainties are 100\% correlated
(one normalization parameter) and the normalization uncertainties are
uncorrelated (five normalization parameters).  The $|F_K|^2$ fit was
the only instance for which the different assumptions about the
correlation of the  normalizations yielded noticeably different fit
results.  In this case, the assumption that the normalizations are
uncorrelated (five normalization parameters) produced a substantially
better fit to the data ($\chi^2/{\rm dof}=48.9/44$) than did the
assumption that they are correlated ($\chi^2/{\rm dof}=73.6/48$).  The
better fit is plotted as a solid line in Fig.~\KPKM.  The fit preferred
a resonance of width 0.17~GeV at mass 1.35 GeV and a second resonance
of width 0.24~GeV at mass 1.68~GeV.

To evaluate the sensitivity of the result to the parameterization, a
second fit was performed with the amplitudes and phases of the $\rho$
and $\omega$ allowed to vary as free parameters.  No appreciable
differences differences from the first pair of fits were observed.
Evaluating equation \dispint\ from $s=4m_{K^+}^2$ to $s=3.24$~GeV$^2$,
we find the $K^+K^-$ contribution to $\dalhadmz$ to be
$$\dalhad^{K^+K^-}(M_Z^2)=0.000356\pm0.000032({\rm
exp})\pm0.000030({\rm param}) \eqn\kpkmans $$
where the parameterization uncertainty reflects the difference obtained
from the two $\chi^2$ definitions.  The two techniques for the
estimation of the experimental uncertainty (discussed in Section~2.1)
yield consistent results in this case.

\section{The Resonances}

The resonances comprise the remaining portion of the total $\ee$ cross
section.    The total cross section for each resonance can be
represented by a relativistic Breit-Wigner form with an
energy-dependent total width
\ref{This form follows from the assumption that the resonance adds an
imaginary part to the photon propagator ${\rm
Im}\Pi_{\gamma\gamma}(s)=-\sqrt{s}\Gamma_{tot}(s)$ and from the
inclusion of final state phase space factors into the cross section.
This form should be a better approximation near threshold which is
included in the range of integration for the $\phi$ and $\omega$
resonances.},
$$\sigma_{res}(s)={12\pi\over m}{\sqrt{s}\Gamma_{ee}\Gamma_{fs}(s)\over
(s-m^2)^2+s\Gamma_{tot}^2(s)}, \eqn\bw      $$
where: $m$, $\Gamma_{ee}$, and $\Gamma_{tot}$ are the mass, electronic
width, and energy-dependent total width of the resonance; and
$\Gamma_{fs}$ is the energy-dependent width corresponding to the final
states considered in the analysis.  Note that the electronic widths are
physical widths (not corrected for vacuum polarization effects).
In order to incorporate the Breit-Wigner cross section described by
equation~\bw\ into equation~\dispint, it must be scaled to the
electromagnetic point cross section,
$\sigma_{\mu\mu}(s)=4\pi\alpha^2(s)/3s$, yielding the following
expression,
$$\dalhad^{res}(q^2)={\alpha_0q^2\over4\pi^2}{\rm
P}\int_{4m_\pi^2}^\infty
ds{\sigma_{res}(s)\over\alpha^2(s)[q^2-s]}, \eqn\sigint $$
which has the slightly unpleasant feature that it incorporates
$\alpha(s)$, the quantity that we are attempting to evaluate, into the
integrand.  To avoid this problem, we use the $\dalhad(s)$
parameterization given in Ref.~\burk\ to generate a first-order
estimate of $\alpha(s)$ for use in equation~\sigint.
Note that equation~\sigint\ is often written with $\alpha(s)$ replaced
by $\alpha_0$.  This is correct only if the cross section
$\sigma_{res}$ is replaced by the tree-level one,
$\sigma_{res}^0=\sigma_{res}\cdot\alpha_0^2/\alpha^2(s)$.  The factor
$\alpha_0^2/\alpha^2(s)$ is often absorbed into equation~\bw\ by
defining the tree-level electronic width
$\Gamma_{ee}^0\equiv\Gamma_{ee}\cdot\alpha_0^2/\alpha^2(m^2)$.

Equation~\sigint\ is evaluated for the $\omega(782)$, $\phi(1020)$,
$\psi$-family, and $\Upsilon$-family resonances by performing a
Simpson's rule integration over the interval $m-60\Gamma_{tot}$ to
$m+60\Gamma_{tot}$ (the lower limit of the $\omega$ integration is the
threshold for $3\pi$ decay).  The energy-dependent total widths of the
$\psi$ and $\Upsilon$ resonances are assumed to scale as $\sqrt{s}$,
$$\Gamma_{tot}(s)={\sqrt{s}\over m}\Gamma_{tot}(m), \eqn\sdepw  $$
where $m$ is the mass of the resonance and $\Gamma_{tot}(m)$ is the
nominal value of the width.  All $\psi$ and $\Upsilon$ final states are
included in the resonance contribution
[$\Gamma_{fs}(s)=\Gamma_{tot}(s)$].  The energy-dependent total width
of the $\phi(1020)$ is given by equation~\gamphi.  The width
$\Gamma_{fs}(s)$ for the $\phi$ is adjusted to exclude the $K^+K^-$
final state (discussed in Section~2.4).  The energy-dependent total
width of the $\omega(782)$ is given by the following expression which
assumes that all final states are $\pi^+\pi^-$, $\pi^0\gamma$, or
$\pi^+\pi^-\pi^0$,
$$\Gamma_\omega(s)=\Gamma_\omega^0\left\lbrace{\sqrt{s}\over
m_\omega}\left[0.022{\beta_\pi^3(s)\over\beta_\pi^3(m_\omega^2)} +
0.085{(1-m_\pi^2/s)^3\over(1-m_\pi^2/m_\omega^2)^3}\right]
+0.893G^\omega_{3\pi}(s) \right\rbrace, \eqn\gamome $$
where: $m_\omega$ is the mass of the $\omega$, $\Gamma_\omega^0$ is the
nominal value\refmark{\pdg} of the $\omega$ width,
$\beta_\pi(s)=\sqrt{1-4m_\pi^2/s}$ is the velocity of the charged pion,
and $G^\omega_{3\pi}(s)$ is a function which is normalized to unity at
$s=m_\omega^2$ and is proportional to the decay rate for
$\omega\to3\pi$ assuming a constant matrix element (phase space
weighting).  The width $\Gamma_{fs}(s)$ for the $\omega$ is adjusted to
exclude the $\pi^+\pi^-$ final states which are included in the
$|F_\pi|^2$ contribution.

The masses and widths used to evaluate equation~\sigint\ are taken from
the 1994 Review of Particle Properties\refmark{\pdg}.  The Particle
Data Group does not apply a consistent set of definitions to the
parameters of all resonances.  The electronic widths of the $\psi$ and
$\Upsilon$ families are defined to be the physical ones and are derived
from fits performed by the PDG itself.  The electronic widths of the
$\omega$ and $\phi$ resonances are determined from measurements of the
total widths and electronic branching fractions $B_{ee}$. In both
cases, the total widths are the correct physical ones.  The average
value of $B_{ee}(\omega)$ is dominated by peak cross section
measurements of the CMD%
\Ref\cmdome{L.M.~Barkov, \etal, {\it JETP Lett.} {\bf 46}, 164 (1987).}
and ND%
\Ref\nd{S.I.~Dolinsky, \etal, {\it Z. Phys.} {\bf C42}, 511 (1989).}
Collaborations which are corrected (partly) for vacuum polarization
effects and lead to a determination of $\Gamma_{ee}^0(\omega)$.  The
case of the $\phi$ is less clear.  Of the three most precise
measurements of $B_{ee}(\phi)$, those of the DM1%
\Ref\dmophi{DM1 Collaboration: G.~Parrour, \etal, {\it Phys. Lett.}
{\bf 63B}, 357 (1976).}
and OLYA%
\Ref\olyaphi{OLYA Collaboration: A.D.~Bukin, \etal, {\it Sov. J. Nucl.
Phys.} {\bf 27}, 516 (1978).}
Collaborations are not corrected for vacuum polarization effects and
lead to a determination of $\Gamma_{ee}(\phi)$.  The most precise
measurement is a later OLYA result which has about the same precision
as the combination of the two preceeding results but is reported in an
unpublished preprint which is no longer available for inspection.  The
result may (or may not) be corrected for vacuum polarization effects.
We make the assumption that the RPP value of $\Gamma_{ee}(\phi)$ is the
physical one.  This assumption cannot be wrong by more than one half of
the total vacuum polarization correction (1.6\%) which we include in
the uncertainty on $\Gamma_{ee}(\phi)$.

The leptonic widths are corrected for incomplete vacuum polarization
correction to the normalizing cross sections (see Appendix~A) before
equation~\sigint\ is evaluated.  The results are listed in Table~1
along with those derived in Sections~2.3 and 2.4.  The experimental
uncertainties are evaluated by assuming that the uncertainties on the
masses, total widths, electronic widths, and relevant branching ratios
are uncorrelated.  The parameterization uncertainties are evaluated by
repeating the calculation with a constant-width, constant-mass
Breit-Wigner cross section.

\section{Final Result}

The various contributions to $\dalhadmz$ are summarized and summed in
Table~1.  The resulting value is
$$\dalhadmz=0.02752\pm0.00046. \eqn\dalans $$
Including the leptonic contribution, we find $\alpha^{-1}(M_Z^2)$ to
be,
$$\alpha^{-1}(M_Z^2)=128.96\pm0.06,  \eqn\answer    $$
where the uncertainties on the lepton masses contribute negligibly to
the total uncertainty.  This result differs by one of its standard
deviations from the (common) result given in References~\ej\ and \bp\
and it differs by 0.3 standard deviations from the result given in
Reference~\mz.  However, since the different analyses make use of many
of the same inputs, the results are not independent measurements of
$\dalhadmz$ but reflect differences in assumptions and technique.
%
\vskip 0.15in
\vbox{
\centerline{\vbox{\hsize=5.2in\singlespace\noindent
{\bf Table 1}: Summary of the various contributions to $\dalhad$.}}
\vskip .1 in
\tabskip=1em plus2em minus.5em
\def\thickfill{\leaders\hrule height1.5pt\hfill}
\vbox{\offinterlineskip
\halign to \hsize
{\vrule width1.5pt#&\strut\hfil#\hfil&\vrule#
&\strut\hfil#\hfil&\vrule#
&\strut\hfil#\hfil&\vrule#
&\strut\hfil#\hfil&\vrule#
&\strut\hfil#\hfil&\vrule width1.5pt#\cr
 \omit\span\omit\span\omit\span\omit\span\omit
 \span\omit\span\omit
 \span\omit\span\omit
 \span\omit\span\omit \thickfill\cr
height 4 pt &\omit&&\omit&&\omit&
            &\omit&&\omit& \cr
& Contribution && W Region (GeV) && $\dalhadmz$ && $\dalexp$ &
& $\dalparam$ &\cr
height 4 pt &\omit&&\omit&&\omit&
            &\omit&&\omit& \cr
 \omit\span\omit\span\omit\span\omit\span\omit
 \span\omit\span\omit
 \span\omit\span\omit
 \span\omit\span\omit \hrulefill\cr
height 4 pt &\omit&&\omit&&\omit&
            &\omit&&\omit& \cr
& Continuum && 1.0-$\infty$ && 0.022106 && 0.000366 && 0.000196 &\cr
height 4 pt &\omit&&\omit&&\omit&
            &\omit&&\omit& \cr
& $\pi^+\pi^-$ && 0.280-2.0 && 0.003240 && 0.000057 && 0.000169 &\cr
height 4 pt &\omit&&\omit&&\omit&
            &\omit&&\omit& \cr
& $K^+K^-$ && 0.987-1.8 && 0.000356 && 0.000032 && 0.000030 &\cr
height 4 pt &\omit&&\omit&&\omit&
            &\omit&&\omit& \cr
& Resonances && $\omega^{(a)}$ && 0.000307 && 0.000010 && 0.000003 &\cr
height 4 pt &\omit&&\omit&&\omit&
            &\omit&&\omit& \cr
& {\tt "} && $\phi^{(b)}$ && 0.000296 && 0.000012 && 0.000004 &\cr
height 4 pt &\omit&&\omit&&\omit&
            &\omit&&\omit& \cr
& {\tt "} && $\psi$ (6 states) && 0.001101 && 0.000059 && 0.000023 &\cr
height 4 pt &\omit&&\omit&&\omit&
            &\omit&&\omit& \cr
& {\tt "} && $\Upsilon$ (6 states) && 0.000118 && 0.000005 && 0.000003
&\cr
height 4 pt &\omit&&\omit&&\omit&
            &\omit&&\omit& \cr
 \omit\span\omit\span\omit\span\omit\span\omit
 \span\omit\span\omit
 \span\omit\span\omit
 \span\omit\span\omit \hrulefill\cr
height 4 pt &\omit&&\omit&&\omit&
            &\omit&&\omit& \cr
& Total &&  && 0.02752 && 0.00038 && 0.00026 &\cr
height 4 pt &\omit&&\omit&&\omit&
            &\omit&&\omit& \cr
 \omit\span\omit\span\omit\span\omit\span\omit
 \span\omit\span\omit
 \span\omit\span\omit
 \span\omit\span\omit \thickfill\cr
}}\noindent
$^{(a)}$Doesn't include $\pi^+\pi^-$ final states.\nextline
$^{(b)}$Doesn't include $K^+K^-$ final states.
}

\section{Detailed Comparison With Reference~\ej}

\Picture\WOCB
\file={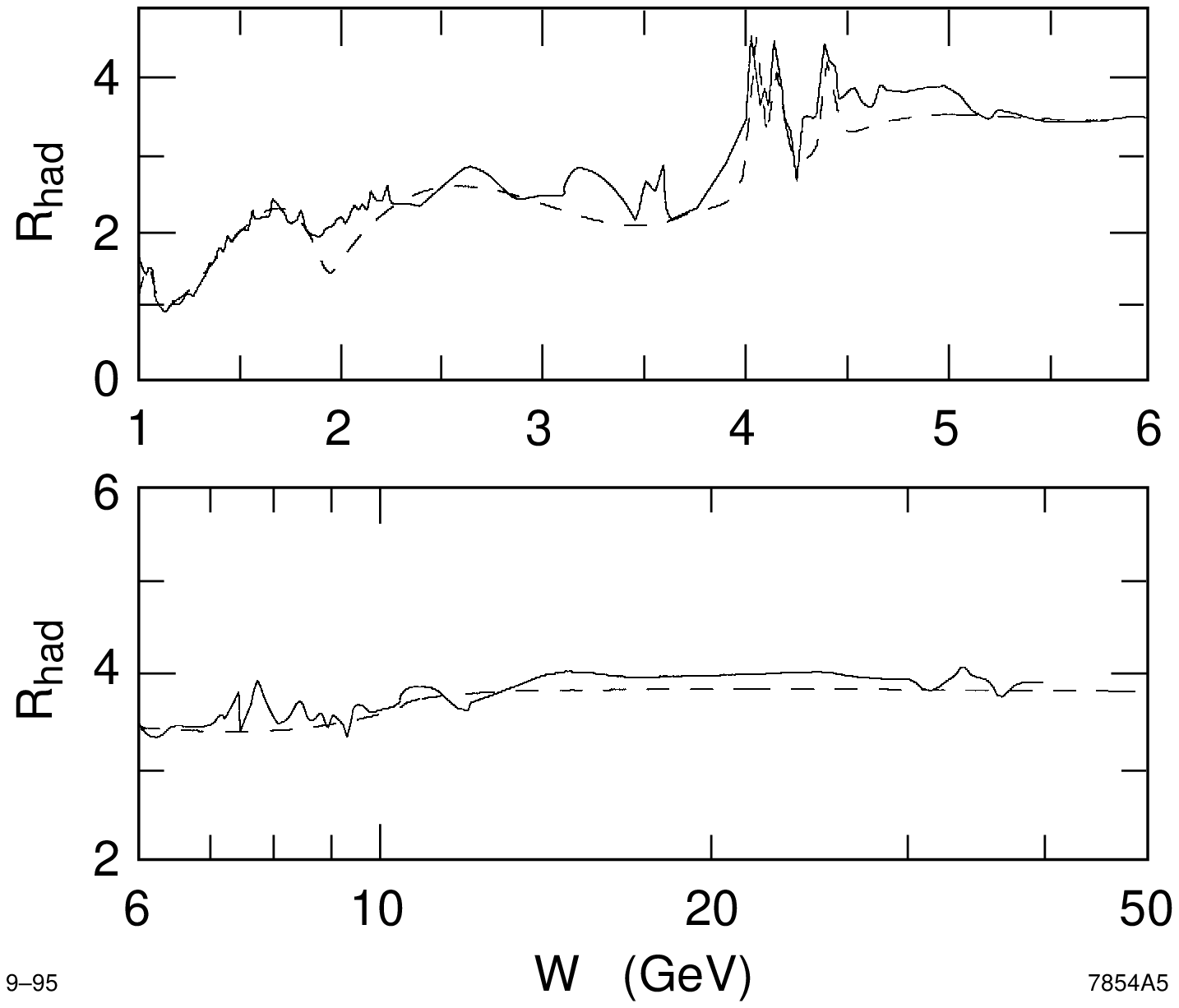}
\height=3in
\width=5in
\caption={\noindent
A comparison of our total $\rhad$ function (dashed curve) before the
inclusion of the Crystal Ball measurement at 3.67~GeV with that from
Reference~\ej.}
\savepicture\piWOCB
\Picture\WCB
\file={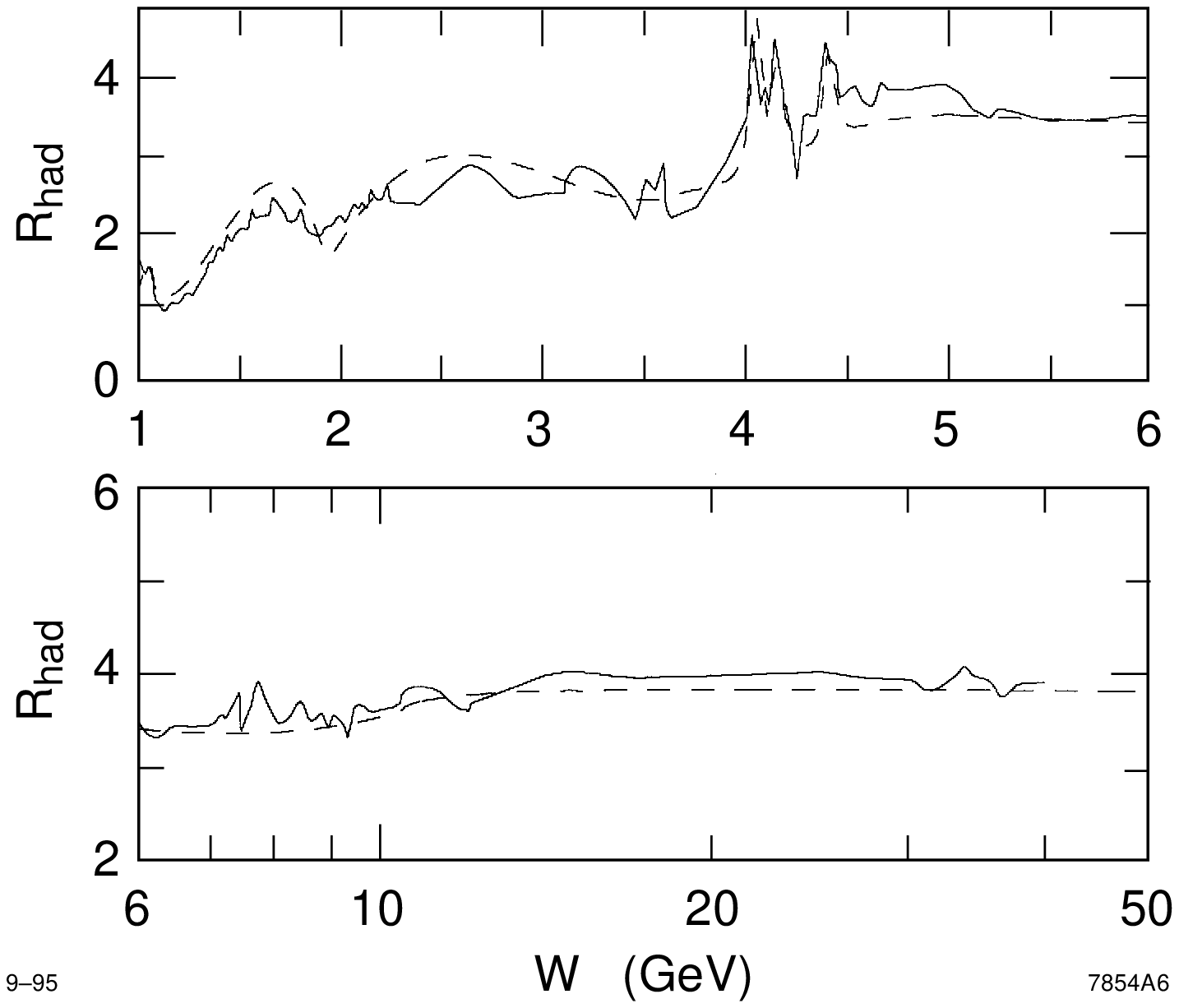}
\height=3in
\width=5in
\caption={\noindent
A comparison of our total $\rhad$ function (dashed curve) after the
inclusion of the Crystal Ball measurement at 3.67~GeV with that from
Reference~\ej.}
\savepicture\piWCB
The result of Eidelman and Jegerlehner\refmark{\ej} (henceforth E-J) is
based almost entirely upon the trapezoidal integration of locally
averaged data points.  Only the narrow resonances are treated
parametrically.  E-J have published their composite compilation of the
function $\rhad(W)$ in a series of figures and include a detailed
breakdown of the contributions of various energy intervals to
$\dalhad(M_Z^2)$.  Since the E-J compilation excludes narrow
resonances, we construct the function $R_{sum}$ to include the same
final states,
$$ R_{sum}(W) = \rfit(W)+{1\over4}|F_\pi(W)|^2\beta_\pi^3
+{1\over4}|F_K(W)|^2\beta_K^3+\sum_{i=1}^5\sigma_{res}^i(W),
\eqn\rsumdef $$
where the sum includes the $\omega(782)$, $\phi(1020)$, $\psi(4040)$,
$\psi(4160)$, and $\psi(4415)$ resonances.  A comparison of their
$\rhad$ compilation ($\rhad^{EJ}$) with $R_{sum}$ in the region
$W=1-50$~GeV is shown in Figures~\WOCB\ and \WCB.  The $\rhad^{EJ}$
compilation is shown as the solid curve in both figures.  The dashed
curve in Figure~\WOCB\ shows $R_{sum}$ before the inclusion of the
Crystal Ball measurement at 3.67~GeV.  The dashed curve in Figure~\WCB\
shows $R_{sum}$ after the inclusion of the new data point.  The peak of
the $\phi$ between 1.00~GeV and 1.04~GeV is suppressed in both figures.

A comparison of Figures~\WOCB\ and \WCB\ shows the effect of the
Crystal Ball measurement quite clearly.  Before the point is added to
the analysis, there is reasonable agreement between the functions
$R_{sum}(W)$ and $\rhad^{EJ}(W)$ in the region 1.0-1.8~GeV.  Between
1.8~GeV and 3.6~GeV, $\rhad^{EJ}$ is generally larger than $R_{sum}$.
After the introduction of the Crystal Ball measurement, the
$\gamma\gamma2$ measurements are renormalized to larger values and the
fitting function generally exceeds the $\rhad^{EJ}$ compilation
throughout the region.

\centerline{\piWOCB}

The agreement between $R_{sum}$ and $\rhad^{EJ}$ in the charm threshold
region between 3.6~GeV and 5.0~GeV is also quite poor.  The $R_{sum}$
function follows the shape of the DASP fit to the continuum under the
$\psi(4040)$, $\psi(4160)$, $\psi(4415)$ resonances and includes the
resonances explicitly for comparison.  The size of the continuum
portion is determined at 3.6~GeV and 5.0~GeV by the most precise data
in those regions (Crystal Ball data in both cases) yielding a
continuous result.  The $\rhad^{EJ}$ compilation generally exceeds
$R_{sum}$ throughout the region reflecting the fact that DASP and PLUTO
generally measured large values of $\rhad$ in the region.  The more
precise Crystal Ball measurements begin at 5.0~GeV and pull the
$\rhad^{EJ}$ function to smaller values, creating an apparent structure
in the 4.4-5.0~GeV region.  The apparent structure is not seen by any
of the experiments that have measured the shape and magnitude of
$\rhad(W)$ in this region and is entirely a consequence of ignoring the
shape information inherent in the data (a more correct procedure would
renormalize the data sets so that the integrated function was smooth
and continuous in the 5-5.2 GeV region).

\centerline{\piWCB}

In the region $W=5-10$~GeV, the agreement of the $R_{sum}$ and
$\rhad^{EJ}$ functions is somewhat better except for some wiggles in
$\rhad^{EJ}$ at the larger energies.  Above b-threshold and below
$W=40$~GeV (where the authors of Reference~\ej\ begin to use
perturbative QCD), the $\rhad^{EJ}$ compilation is somewhat larger than
$R_{sum}$ reflecting the fact the the PEP/PETRA measurements of $\rhad$
are somewhat larger than those predicted by perturbative QCD with
currently favored values of $\alpha_s(M_Z^2)$.

The differences shown qualitatively in Figure~\WCB\ are quantified in
Table~2 using the detailed breakdown scheme presented in Reference~\ej.
The entries in square brackets are from Ref.~\ej\ before the
application of corrections for incomplete vacuum polarization
correction.  In Appendix~A, we demonstrate that the non-application of
this correction is generally a more accurate approximation than the use
of the factor favored by the authors of Ref.~\ej.  A comparison of our
result with the bracketed quantities (or the means of the pairs of
quantities) is probably the more relevant one.    Note that our value
for the contribution of $\pi^+\pi^-$ final states in the interval
$W=0.28-0.81$~GeV of 24.11 (in units of 10$^{-4}$) is somewhat smaller
than the value of 26.08 given in Reference~\ej.  The difference may be
due in part to the preference of our fit for smaller values of
$|F_\pi|^2$ than the central values of the OLYA measurements between
0.6~GeV and 1.0~GeV (see Fig.~\PIPI).  The opposite behavior is observed
when the full function used in the analysis of Kinoshita, Nizic, and
Okamoto\refmark{\kno} is fit to the data.  The large-energy tail of this
function decreases with energy more steeply than do the data points.
A fit to this function prefers larger values of $|F_\pi|^2$ than the
central values of the OLYA measurements between 0.6~GeV and 1.0~GeV
yielding a contribution to $\dalhadmz$ of 25.39.  Excluding the influence
of the steeply falling tail by restricting the fit of the KNO function to
the region $W<1.0$~GeV relaxes some of the bias and yields a $\dalhadmz$
contribution of 24.76.  These differences are reflected in the large size
of the parameterization uncertainty given in Table~1.

We conclude that the agreement of our analysis with one based almost
entirely on trapezoidal integration is somewhat poorer than a
comparison of the final $\dalhad(M_Z^2)$ results would indicate.  Part
of the discrepancy is caused by the loss of shape information from
multi-point measurements inherent in the averaging procedure which
treats the individual measurements as independent.  An associated side
effect is that sparse, newer measurements influence the integrated
function only over an interval between neighboring older measurements.
The addition of the precise Crystal Ball point (which fixes the
normalization of $\rhad$ over a large region in our analysis) to a
trapezoidal analysis would affect only a very small region.
Conversely, the trapezoidal analysis remains influenced by older
measurements until they are replaced by newer measurements at the same
or very nearby energies.  The effect of the apparent structure in the
charm threshold region or the large $\rhad$ values from the PEP/PETRA
region will persist until replaced (or influenced) by newer
measurements at the same energies.  The use of a continuous fitting
function in our analysis allows us to interpolate between sparse but
precise points.  For these reasons, we do indeed ``believe more in the
integration of our fits than in trapezoidal integration'' as noted by
the authors of Ref.~\ej.
%
\vskip 0.15in
\vbox{
\centerline{\vbox{\hsize=5.2in\singlespace\noindent
{\bf Table 2}: Comparison of the various contributions to
$\dalhad(M_Z^2)$ with those published in Ref.~\ej (in units of
10$^{-4}$).  The entries in square brackets are from Ref.~\ej\ before
the application of corrections for incomplete vacuum polarization
correction.}}
\vskip .10 in
\tabskip=1em plus2em minus.5em
\def\thickfill{\leaders\hrule height1.5pt\hfill}
\vbox{\offinterlineskip
\halign to \hsize
{\vrule width1.5pt#&\strut\hfil#\hfil&\vrule#&\strut\hfil#\hfil&\vrule#
&\strut\hfil#\hfil&\vrule#&\strut\hfil#\hfil&\vrule width1.5pt#\cr
 \omit\span\omit\span\omit\span\omit\span\omit
 \span\omit\span\omit\span\omit\span\omit \thickfill\cr
height 4 pt &\omit&&\omit&
            &\omit&&\omit& \cr
& Final State && $W$ Interval (GeV) && This Work &
& Reference~\ej &\cr
height 4 pt &\omit&&\omit&
            &\omit&&\omit& \cr
 \omit\span\omit\span\omit\span\omit\span\omit
 \span\omit\span\omit\span\omit\span\omit \hrulefill\cr
height 4 pt &\omit&&\omit&
            &\omit&&\omit& \cr
& $\rho$ && 0.28-0.81 && 24.11 && 26.08 [26.23] &\cr
height 4 pt &\omit&&\omit&
            &\omit&&\omit& \cr
& $\omega$ && 0.42-0.81 && 2.87 && 2.93 [2.96] &\cr
height 4 pt &\omit&&\omit&
            &\omit&&\omit& \cr
& $\phi$ && 1.00-1.04 && 5.03 && 5.08 [5.15] &\cr
height 4 pt &\omit&&\omit&
            &\omit&&\omit& \cr
& ${\rm J}/\psi$ &&  && 11.01 && 11.34 [11.93] &\cr
height 4 pt &\omit&&\omit&
            &\omit&&\omit& \cr
& $\Upsilon$ &&  && 1.18 && 1.18 [1.27] &\cr
height 4 pt &\omit&&\omit&
            &\omit&&\omit& \cr
& all hadrons && 0.81-1.40 && 13.55 && 13.83 [13.99] &\cr
height 4 pt &\omit&&\omit&
            &\omit&&\omit& \cr
& all hadrons && 1.40-3.10 && 30.42 && 27.62 [28.23] &\cr
height 4 pt &\omit&&\omit&
            &\omit&&\omit& \cr
& all hadrons && 3.10-3.60 && 5.62 && 5.82 [5.98] &\cr
height 4 pt &\omit&&\omit&
            &\omit&&\omit& \cr
& all hadrons && 3.60-9.46 && 48.16 && 50.60 [50.50] &\cr
height 4 pt &\omit&&\omit&
            &\omit&&\omit& \cr
& all hadrons && 9.46-40.0 && 90.67 && 93.07 &\cr
height 4 pt &\omit&&\omit&
            &\omit&&\omit& \cr
& all hadrons && 40.0-$\infty$ && 42.64 && 42.82 &\cr
height 4 pt &\omit&&\omit&
            &\omit&&\omit& \cr
 \omit\span\omit\span\omit\span\omit\span\omit
 \span\omit\span\omit\span\omit\span\omit \hrulefill\cr
height 4 pt &\omit&&\omit&
            &\omit&&\omit& \cr
& Total &&  && 275.2 && 280.4 [282.1] &\cr
height 4 pt &\omit&&\omit&
            &\omit&&\omit& \cr
 \omit\span\omit\span\omit\span\omit\span\omit
 \span\omit\span\omit\span\omit\span\omit \thickfill\cr
}}}

\chapter{Conclusions}

We have reevaluated the hadronic part of the electromagnetic vacuum
expectation value using the standard dispersion integral approach that
utilizes the hadronic cross section measured in $\ee$ experiments as
input.  Previous analyses are based upon point-by-point trapezoidal
integration which does not treat experimental errors in an optimal way.
We use a technique that weights the experimental inputs by their stated
uncertainties, includes correlations, and incorporates some
refinements.  We find the five-flavor hadronic contribution to the
fractional change in the electromagnetic coupling constant at
$q^2=M_Z^2$, $\dalphamz$, to be $0.02752\pm0.00046$,
which leads to a value of the electromagnetic coupling constant,
$\alpha^{-1}(M_Z^2) = 128.96\pm0.06$.

The current generation of $Z$-pole asymmetry measurements have already
determined the effective weak mixing angle $\swein$ to a precision of
$\pm0.00028$%
\ref{See the presentation of A.~Olshevski at the 1995 International
Europhysics Conference on High Energy Physics, Brussels, Belgium.}.
Future measurements may improve the determination to the $\pm0.00020$
level.  This is comparable to the theoretical uncertainty of
$\pm$0.00016 which follows from the $\pm$0.06 uncertainty on
$\alpha^{-1}(M_Z^2)$.  It is clear that improved understanding of
$\alpha(M_Z^2)$ is desirable and it is also clear (from Figure~\INTUNC)
that improved understanding requires improved data in the $W=1-5$~GeV
region.  Additionally, the differences with the trapezoidal approach
noted in Section~2.7 stem from questions dealing with the optimal use
of rather poor quality data.  Improved data will tend to make these
issues less important.  Among the active experimental programs of the
world, only the BES Collaboration at the Beijing Electron Positron
Collider is positioned to make improved measurements of $\rhad$ in the
region $W=2-5$~GeV.  They are urged to include them in their long term
planning.

\endpage
\ack

This work was originally motivated by the Long Term Planning Study
organized by the Division of Particles and Fields of the American
Physical Society.  The author would like to thank Michael Peskin for
his useful discussions, technical advice, and comments on this
manuscript.  The author would also like to thank Tatsu Takeuchi and
Bill Lockman for pointing out the existence of References~\dagostini\
and \cbcthr, respectively.
This work was supported by Department of Energy Contract No.
DE-AC03-76SF00515.


\chapter{Appendix A: Vacuum Polarization Corrections}

\section{Corrections to $\rhad$}

The quantity $\rhad$ is the ratio of s-channel cross sections and can
be written as follows,
$$\rhad\equiv{\sigma_{had}(s)\over\sigma_{\mu\mu}(s)}={\sigma_{had}^0(s)
\over
\sigma_{\mu\mu}^0(s)}, \eqn\rdefs $$
where the tree-level cross sections $\sigma^0(s)$ are related to the
physical ones (already corrected for initial state radiation) by the
simple expression,
$\sigma^0(s)=\sigma(s)\alpha_0^2/\alpha^2(s)$.  Since radiative
corrections calculations combine external photonic corrections and
virtual corrections, it is more straightforward for experiments to
extract $\sigma_{had}^0(s)$ from their data than it is to extract
$\sigma_{had}(s)$.  Note that $\sigma_{\mu\mu}^0(s)$ is a simple
numerical constant which is applied to the measured cross section after
radiative corrections.

In Reference~\ej, Eidelman and Jegerlehner point out that many of the
earlier measurements of $\rhad$, $|F_\pi|^2$, and $|F_K|^2$ were
corrected for leptonic vacuum polarization effects but were not
corrected for hadronic vacuum polarization effects.  To rectify this
problem, they make the assumption that individual experiments directly
measure hadronic cross sections and apply the factor,
$$r_c^{EJ} =
\left[1+2\dalpha_\ell(s)\right]{\alpha_0^2\over\alpha^2(s)},
\eqn\ejcorr $$
to all measurements of $\rhad$, $|F_\pi|^2$, and $|F_K|^2$ below the
$J\psi$ and to the Mark~I measurements below charm threshold.

Unfortunately, the integrated luminosity for each measurement must be
determined from the measurement of an additional physical process.
Thus, experiments rarely measure cross sections directly but nearly
always measure {\it the ratios of cross sections}.  In this case, the
measured value of $\rhad$ (or $|F|^2\beta^3/4$) is determined from the
ratio of the number of observed hadronic events $N_{had}$ to the number
of observed normalizing events $N_{norm}$,
$$\rhad={N_{had}(1+\delta_{rc})\over N_{norm}\varepsilon}\cdot
{\sigma_{norm}(s)\over\sigma_{\mu\mu}^0(s)}, \eqn\rmeas  $$
where $\delta_{rc}$ incorporates all radiative corrections to the
hadronic yield, $\varepsilon$ is the efficiency-acceptance product for
hadronic events, and $\sigma_{norm}$ is the physical cross section for
the normalizing events (including all radiative effects) integrated
over the acceptance used for the luminosity measurement.  We note that
the incomplete application of vacuum polarization corrections is a
problem that applies to both the hadronic and normalizing cross
sections.  In this case, the actual correction should be
$$r_c = {\alpha_\ell^2(s)\over\alpha^2(s)}\cdot
{\sigma_{norm}(s)\over\sigma_{norm}^\ell(s)} \simeq
{\alpha^2_0\left[1+2\dalpha_\ell(s)\right]\over\alpha^2(s)}\cdot
{\sigma_{norm}(s)\over\sigma_{norm}^\ell(s)}, \eqn\corcorr    $$
where $\alpha_\ell(s)$ and $\sigma_{norm}^\ell(s)$ incorporate leptonic
vacuum polarization corrections only.  The difference between the two
right-hand terms involves the (numerically insignificant) question of
whether the original vacuum polarization corrections were performed to
all orders or to first order only.

All of the early measurements of $\rhad$, $|F_\pi|^2$, and $|F_K|^2$
are normalized to the number of lepton pairs observed in some portion
of each apparatus.  Most of the experiments did not have (or did not
use) small-angle Bhabha scattering luminosity monitors but relied
instead upon large-angle lepton pairs observed in the central region of
each detector.  The combination of the leptonic final states and
geometric acceptance used by the major experiments is summarized in
Table~3.  Several experiments use muon pairs to normalize their
results.  Since the vacuum polarization corrections to s-channel
processes can be factorized (see equation \rdefs), the correction
factor given by equation \corcorr\ is identically 1.  The remaining
experiments use a combination of $e^+e^-$ and $\mu^+\mu^-$ events or
$e^+e^-$ events alone to normalize their results.  The electron-pair
final states are produced by the sum of s- and t-channel subprocesses.
The vacuum polarization corrections to the dominant t-channel
contributions are proportional to $\alpha^2(-t)$.  Since the t-channel
contribution dominates the Bhabha cross section, the correction factor
$r_c$ is given roughly by the following expression,
$$ r_c\sim
{\alpha^2(-t)\over\alpha^2(s)}\cdot{\alpha_\ell^2(s)\over\alpha_\ell^2(-
t)}. \eqn\appcor $$
The key point in this discussion is that the dependence of
$\alpha(q^2)$ upon the scale $q^2$ is logarithmic and the magnitude of
$-t$ at the large angles used by most of the experiments is comparable
to $s$ (typically, $-t/s=0.2\to0.4$).  For this $-t$ range, the first
ratio in equation~\appcor\ is typically a few percent less than unity
and the second ratio is a few percent larger than unity.
The net correction is therefore quite small.  A complete
calculation of the correction factor $r_c$ for requires that all
luminosity event selection criteria be incorporated into complete
calculations of $\sigma_{norm}$ and $\sigma_{norm}^\ell$ (incorporating
all radiative corrections).  Rather than undertake such an arduous
procedure, we estimate the size of the correction from a simplified
calculation which accounts for vacuum polarization effects and
approximate angular acceptance.  The estimate uses the low energy
parameterization of $\dalhad$ found in Ref.~\burk.  The results of this
estimate are listed in Table~3 along with the correction advocated by
the authors of Ref.~\ej.  Note that the corrections to the pseudoscalar
form factors are estimated assuming that the original leptonic vacuum
polarization corrections included electron and muon contributions.  The
corrections to the $\rhad$ measurements are estimated assuming that the
original corrections included only the electron contribution.

\vskip 0.15in
\vbox{
\centerline{\vbox{\hsize=5.2in\singlespace\noindent
{\bf Table 3}: Summary of the incomplete vacuum polarization correction
factor $r_c$ and that of Ref.~\ej, $r_c^{EJ}$.}}
\vskip .1 in
\tabskip=1em plus2em minus.5em
\def\thickfill{\leaders\hrule height1.5pt\hfill}
\vbox{\offinterlineskip
\halign to \hsize
{\vrule width1.5pt#&\strut\hfil#\hfil&\vrule#
&\strut\hfil#\hfil&\vrule#
&\strut\hfil#\hfil&\vrule#
&\strut\hfil#\hfil&\vrule#
&\strut\hfil#\hfil&\vrule#
&\strut\hfil#\hfil&\vrule#
&\strut\hfil#\hfil&\vrule width1.5pt#\cr
 \omit\span\omit\span\omit\span\omit\span\omit
 \span\omit\span\omit
 \span\omit\span\omit
 \span\omit\span\omit
 \span\omit\span\omit
 \span\omit\span\omit \thickfill\cr
height 4 pt &\omit&&\omit&&\omit&&\omit&
            &\omit &&\omit&&\omit& \cr
& Exp. && Meas. && Norm. && $|\cos(\theta)|$ && $W$ (GeV) &
& $r_c$ && $r_c^{EJ}$ &\cr
height 4 pt &\omit&&\omit&&\omit&&\omit&
            &\omit &&\omit&&\omit& \cr
 \omit\span\omit\span\omit\span\omit\span\omit
 \span\omit\span\omit
 \span\omit\span\omit
 \span\omit\span\omit
 \span\omit\span\omit
\span\omit\span\omit \hrulefill\cr
height 4 pt &\omit&&\omit&&\omit&&\omit&
            &\omit &&\omit&&\omit& \cr
& NA7\refmark{\nas} && $|F_\pi|^2$ && $\mu\mu$ &&
$-$.875$\to$.997$^{(a)}$ && 0.320 && 1.0000 && 0.9982 &\cr
&  &&  && &&  && 0.422 && 1.0000 && 0.9972 &\cr
height 4 pt &\omit&&\omit&&\omit&&\omit&
            &\omit &&\omit&&\omit& \cr
& OLYA\refmark{\vepptwopi,\olya} && $|F_\pi|^2$, $|F_K|^2$ &&
$ee+\mu\mu$ && $<$0.71 && 0.400 && 0.9984 && 0.9974 &\cr
&  &&  && &&  && 1.397 && 0.9952 && 0.9893 &\cr
height 4 pt &\omit&&\omit&&\omit&&\omit&
            &\omit &&\omit&&\omit& \cr
& CMD\refmark{\vepptwopi,\cmd} && $|F_\pi|^2$, $|F_K|^2$ && $ee+\mu\mu$
&& $<$0.60 && 0.360 && 0.9988 && 0.9978 &\cr
&  &&  && &&  && 0.820 && 0.9970 && 0.9934 &\cr
height 4 pt &\omit&&\omit&&\omit&&\omit&
            &\omit &&\omit&&\omit& \cr
& TOF\refmark{\tof} && $|F_\pi|^2$ && $ee+\mu\mu$ && $<$0.24 && 0.400
&& 0.9990 && 0.9974 &\cr
&  &&  && &&  && 0.460 && 0.9988 && 0.9968 &\cr
height 4 pt &\omit&&\omit&&\omit&&\omit&
            &\omit &&\omit&&\omit& \cr
& $\mu\pi$\refmark{\mupi} && $|F_\pi|^2$ && $ee$ && $<$0.61 && 1.250 &&
0.9958 && 0.9902 &\cr
&  &&  && &&  && 1.520 && 0.9955 && 0.9886 &\cr
height 4 pt &\omit&&\omit&&\omit&&\omit&
            &\omit &&\omit&&\omit& \cr
& MEA\refmark{\mea} && $|F_\pi|^2$, $|F_K|^2$ && $ee$ && $<$0.77 && 1.6
&& 0.9941 && 0.9826 &\cr
&  &&  && $\mu\mu$ &&  && 1.43 && 1.0000 && 0.9838 &\cr
height 4 pt &\omit&&\omit&&\omit&&\omit&
            &\omit &&\omit&&\omit& \cr
& DM1\refmark{\dmopi,\dmok} && $|F_\pi|^2$, $|F_K|^2$ && $ee$ &&
$<$0.50 && 0.480 && 0.9983 && 0.9966 &\cr
&  &&  &&  &&  && 2.060 && 0.9960 && 0.9860 &\cr
height 4 pt &\omit&&\omit&&\omit&&\omit&
            &\omit &&\omit&&\omit& \cr
& DM2\refmark{\dmtpi,\dmtk} && $|F_\pi|^2$, $|F_K|^2$ && $\mu\mu$ &&
$<$0.87 && 1.350 && 1.0000 && 0.9896 &\cr
&  &&  &&  &&  && 2.400 && 1.0000 && 0.9848 &\cr
height 4 pt &\omit&&\omit&&\omit&&\omit&
            &\omit &&\omit&&\omit& \cr
& $\gamma\gamma2$\refmark{\bacci} && $\rhad$ && $ee$ && $<$0.64 && 1.42
&& 0.9933 && 0.9839 &\cr
&  &&  &&  &&  && 3.09 && 0.9935 && 0.9757 &\cr
height 4 pt &\omit&&\omit&&\omit&&\omit&
            &\omit &&\omit&&\omit& \cr
& Mark~I\refmark{\mki} && $\rhad$ && $ee$ && $<$0.60 && 2.60 && 0.9936
&& 0.9772 &\cr
&  &&  &&  &&  && 3.65 && 0.9958 && 0.9756 &\cr
height 4 pt &\omit&&\omit&&\omit&&\omit&
            &\omit &&\omit&&\omit& \cr
& DASP\refmark{\dasp} && $\rhad$ && $ee$ && $<$0.71$^{(b)}$ && 3.6 &&
0.9946 && 1.0000 &\cr
height 4 pt &\omit&&\omit&&\omit&&\omit&
            &\omit &&\omit&&\omit& \cr
& PLUTO\refmark{\pluto} && $\rhad$ && $ee$ && 0.9816$\to$0.9977 && 3.6
&& 0.9756 && 1.0000 &\cr
height 4 pt &\omit&&\omit&&\omit&&\omit&
            &\omit &&\omit&&\omit& \cr
& CMD\refmark{\cmdome} && $\Gamma_{ee}^0(\omega)$ && $ee+\mu\mu$ &&
$<$0.60 && 0.782 && 0.9971 && 0.9904 &\cr
height 4 pt &\omit&&\omit&&\omit&&\omit&
            &\omit &&\omit&&\omit& \cr
& ND\refmark{\nd} && $\Gamma_{ee}^0(\omega)$ && $ee$ && $<$0.65 &&
0.782 && 0.9942 && 0.9904 &\cr
height 4 pt &\omit&&\omit&&\omit&&\omit&
            &\omit &&\omit&&\omit& \cr
\omit\span\omit\span\omit\span\omit\span\omit
 \span\omit\span\omit
 \span\omit\span\omit
 \span\omit\span\omit
 \span\omit\span\omit
\span\omit\span\omit \thickfill\cr
}}\noindent
$^{(a)}$Interval in $\cos\theta$.\nextline
$^{(b)}$Used small-angle $\ee$ events normalized to this large angle
region.
}

The reader should note several things.  The corrections to the $\rhad$,
$|F_\pi|^2$, and $|F_K|^2$ measurements are always a factor of seven or
more smaller than systematic normalization uncertainties associated with
the measurements.  In all cases, the correction applied by the authors
of Ref.~\ej\ overestimates the true size of the correction.  This
overestimate is small where the correction is small but becomes
significant at larger energies where the Eidelman-Jegerlehner correction
exceeds 1\%.  In this region, the non-application of the correction
($r_c=1.0$) is a better approximation than the one used by the authors
of Ref.~\ej.  The Eidelman-Jegerlehner analysis did not correct the
hadronic continuum measurements of the DASP and PLUTO Collaborations
at charm threshold although it appears that neither group applied
hadronic vacuum polarization corrections%
\ref{The original work of Berends and Komen \refmark{\old} in 1976 was
not integrated into a radiative corrections code until the work of
F.A.~Berends and R.~Kleiss, {\it Nucl. Phys.} {\bf B178}, 141 (1981).
The 1978 DASP measurement\refmark{\dasp} explicitly states that the
radiative corrections of G.~Bonneau and F.~Martin, {\it Nucl. Phys.},
{\bf B27}, 381 (1971) were applied.  The 1977 PLUTO measurement
\refmark{\pluto} merely states that radiative corrections were applied.}.
The normalization DASP measurements was determined from the total number
of large-angle Bhabha scattering events and is subject to a small
correction.  The PLUTO experiment normalized its measurements with a
small-angle luminosity monitor which sampled a region of small $-t$.
The cancellation of the vacuum polarization corrections is
correspondingly smaller and the correction is larger.

\section{Corrections to Resonance Parameters}

The Breit-Wigner cross section used in Section~2.5 to calculate the
resonant contribution to $\dalhadmz$ requires the mass, total width,
and electronic width of each resonance as input.  The electronic widths
$\Gamma_{ee}$ are defined to be physical quantities (not corrected for
vacuum polarization effects) and differ from the tree-level quantities
$\Gamma_{ee}^0$ that have been used often in the past.  The electronic
widths for narrow and broad resonances are determined by different
techniques but are always proportional to the peak hadronic cross
section of the resonance (measured in $\ee$ collisions) or to the
measured energy-integral of the hadronic cross section (taken over the
resonance),
$$\Gamma_{ee}
\propto{N_{had}(1+\delta_{rc}^\prime)\over N_{norm}\varepsilon}\cdot
\sigma_{norm}(s), \eqn\gmeas  $$
where all quantities are defined in equation~\rmeas\ except for
$\delta_{rc}^\prime$ which accounts for radiative corrections to the
hadronic yield but excludes vacuum polarization corrections.  The
inclusion of vaccum polarization corrections into $\delta_{rc}^\prime$
($\delta_{rc}^\prime\to\delta_{rc}$) yields a measurement of the
tree-level quantity $\Gamma_{ee}^0$.

As in the case of the cross section and form factor measurements, many
of the older measurements of the electronic widths were not corrected
for hadronic vacuum polarization effects.  It is clear that
measurements of $\Gamma_{ee}^0$ must be corrected by the same
correction factor $r_c$ defined in equation~\corcorr.  However, for
measurements of $\Gamma_{ee}$, vacuum polarization corrections to the
hadronic yield are not applied and the appropriate correction factor
$g_c$ pertains to the normalizing cross section only,
$$g_c = {\sigma_{norm}(s)\over\sigma_{norm}^\ell(s)}. \eqn\eewcorr $$

As was discussed in Section~2.5, the Review of Particle Properties
lists physical widths for the $\psi$- and $\Upsilon$-family resonances
as derived from their own fitting procedure.  The electronic width of
the $\phi(1020)$ is either the physical value or an average of the
tree-level and physical values and is assumed to the the physical one.
The oldest measurements of these quantities were corrected for electron
vacuum polarization effects only and require the application of the
additional correction factor $g_c$.  Estimates of this factor are
listed in Table~4 for measurements of the $\phi$, $J/\psi(1S)$, and
$J/\psi(2S)$ electronic widths.  The weighted average of the $\phi$
correction factors is applied to the PDG value of $\Gamma_{ee}(\phi)$.
The corrections to the $\psi$-family measurements are quite small if
the original measurement was normalized to small-angle Bhabha
scattering and can be as large as 2\% if the large angle cross section
was used as a normalization.  Unfortunately, since the quoted
electronic widths are derived from global fits, it is difficult to
estimate the effect on the final value of $\Gamma_{ee}$.  Therefore, we
do not apply any corrections to the electronic widths of the
$\psi$-family but we do inflate the uncertainties on $\Gamma_{ee}$ by
the size of the largest correction.

Unlike the other resonances, the electronic width of the $\omega(782)$
listed in the Review of Particle Properties is the tree-level one.  We
therefore apply the weighted average of the correction factors $r_c$
listed in Table~3 for the dominant CMD and ND measurements.

\REF\mkipsi{Mark~I Collaboration: A.M.~Boyarski, \etal, {\it Phys. Rev.
Lett.} {\bf 34}, 1357 (1975).}
\REF\ggtpsi{$\gamma\gamma2$ Collaboration: R.~Baldini-Celio, \etal,
{\it Phys. Lett.} {\bf 58B}, 471 (1975).}
\REF\meapsi{MEA Collaboration: B~Esposito, \etal, {\it Nuovo Cimento
Lett.} {\bf 14}, 73 (1975).}
\REF\dasppsi{DASP Collaboration: R.~Brandelik, \etal, {\it Z. Phys.}
{\bf C1}, 233 (1979).}
\REF\mkipsip{Mark~I Collaboration: V.~L\" uth, \etal, {\it Phys. Rev.
Lett.} {\bf 35}, 1124 (1975).}

\vskip 0.2in
\vbox{
\centerline{\vbox{\hsize=5.2in\singlespace\noindent
{\bf Table 4}: Summary of the incomplete vacuum polarization correction
factor $g_c$.}}
\vskip .1 in
\tabskip=1em plus2em minus.5em
\def\thickfill{\leaders\hrule height1.5pt\hfill}
\vbox{\offinterlineskip
\halign to \hsize
{\vrule width1.5pt#&\strut\hfil#\hfil&\vrule#
&\strut\hfil#\hfil&\vrule#
&\strut\hfil#\hfil&\vrule#
&\strut\hfil#\hfil&\vrule#
&\strut\hfil#\hfil&\vrule width1.5pt#\cr
 \omit\span\omit\span\omit\span\omit\span\omit
 \span\omit\span\omit
 \span\omit\span\omit
 \span\omit\span\omit \thickfill\cr
height 4 pt &\omit&&\omit&
            &\omit &&\omit&&\omit& \cr
& Exp. && Res. && Norm. && $|\cos(\theta)|$ && $g_c$ &\cr
height 4 pt &\omit&&\omit&
            &\omit &&\omit&&\omit& \cr
 \omit\span\omit\span\omit\span\omit\span\omit
 \span\omit\span\omit
 \span\omit\span\omit
\span\omit\span\omit \hrulefill\cr
height 4 pt &\omit&&\omit&
            &\omit &&\omit&&\omit& \cr
& DM1\refmark{\dmophi} && $\phi(1020)$ && $ee$ && $<$0.50 && 1.0071
&\cr
height 4 pt &\omit&&\omit&
            &\omit &&\omit&&\omit& \cr
& OLYA\refmark{\olyaphi} && $\phi(1020)$ && $ee$ && $<$0.71 && 1.0052
&\cr
height 4 pt &\omit&&\omit&
            &\omit &&\omit&&\omit& \cr
& Mark~I\refmark{\mkipsi} && $J/\psi(1S)$ && $ee$ && 0.9997$\to$0.9999
&& 1.0000 &\cr
height 4 pt &\omit&&\omit&
            &\omit &&\omit&&\omit& \cr
& $\gamma\gamma2$\refmark{\ggtpsi} && $J/\psi(1S)$ && $ee$ &&
0.9945$\to$0.9986 && 1.0002 &\cr
height 4 pt &\omit&&\omit&
            &\omit &&\omit&&\omit& \cr
& MEA\refmark{\meapsi} && $J/\psi(1S)$ && $ee$ && $<$0.77 && 1.0158
&\cr
height 4 pt &\omit&&\omit&
            &\omit &&\omit&&\omit& \cr
& DASP\refmark{\dasppsi} && $J/\psi(1S)$ && $ee$ && $<$0.71 && 1.0169
&\cr
height 4 pt &\omit&&\omit&
            &\omit &&\omit&&\omit& \cr
& Mark~I\refmark{\mkipsip} && $J/\psi(2S)$ && $ee$ && $<$0.60 && 1.0204
&\cr
height 4 pt &\omit&&\omit&
            &\omit &&\omit&&\omit& \cr
& DASP\refmark{\dasppsi} && $J/\psi(2S)$ && $ee$ && $<$0.71 && 1.0189
&\cr
height 4 pt &\omit&&\omit&
            &\omit &&\omit&&\omit& \cr
\omit\span\omit\span\omit\span\omit\span\omit
 \span\omit\span\omit
 \span\omit\span\omit
\span\omit\span\omit \thickfill\cr
}}}


\refout
\endpage


\bye